\documentclass[12pt,preprint]{aastex}
\usepackage{natbib}
\def\hal{H$\alpha$}

\def\be{\begin{equation}}
\def\ee{\end{equation}}

\def\m{~$\mu$m}
\def\R25{$R_{25}$}

\def\Spitzer{{\it Spitzer}}
\def\GALEX{{\it GALEX}}
\def\PS   {{\it Pan-STARRS1}}
\def\NOAO {{\it NOAO}}
\def\B    {{\it B}}
\def\V    {{\it V}}
\def\Rc   {{\it R$_{\rm C}$}}
\def\Ic   {{\it I$_{\rm C}$}}
\def\u    {{\it u}}
\def\g    {{\it g}}
\def\r    {{\it r}}
\def\i    {{\it i}}
\def\z    {{\it z}}

\def\gps  {{\it g$_{\rm PS1}$}}
\def\rps  {{\it r$_{\rm PS1}$}}
\def\ips  {{\it i$_{\rm PS1}$}}
\def\zps  {{\it z$_{\rm PS1}$}}
\def\J    {{\it J}}
\def\H    {{\it H}}
\def\Ks   {{\it K$_{\rm s}$}}
\def\GALEX{{\it GALEX}}
\def\Spitzer{{\it Spitzer}}
\def\Planck{{\it Planck}}
\def\Herschel{{\it Herschel}}
\def\WISE{{\it WISE}}
\def\2MASS{{\it 2MASS}}

\def\SDSS{{\it SDSS}}
\def\VLA{{\it VLA}}
\def\JCMT{{\it JCMT}}

\begin {document}
\title{UPDATED 34-BAND PHOTOMETRY FOR THE SINGS/KINGFISH SAMPLES OF NEARBY GALAXIES}

\author {
D.~A. Dale\altaffilmark{1},
D.~O. Cook\altaffilmark{2},
H. Roussel\altaffilmark{3},  
J.~A. Turner\altaffilmark{1}, 
L. Armus\altaffilmark{4},
A.~D. Bolatto\altaffilmark{5},
M. Boquien\altaffilmark{6},
M.~J.~I. Brown\altaffilmark{7},
D. Calzetti\altaffilmark{8},
I. De Looze\altaffilmark{9},
M. Galametz\altaffilmark{10,11},
K.~D. Gordon\altaffilmark{12},
B.~A. Groves\altaffilmark{13},
T.~H. Jarrett\altaffilmark{14}
G. Helou\altaffilmark{4},
R. Herrera-Camus\altaffilmark{15},
J.~L. Hinz\altaffilmark{16},
L.~K. Hunt\altaffilmark{17},
R.~C. Kennicutt\altaffilmark{18},
E.~J. Murphy\altaffilmark{19}, 
A. Rest\altaffilmark{12},
K.~M. Sandstrom\altaffilmark{20},  
J.-D.~T. Smith\altaffilmark{21}, 
F.~S. Tabatabaei\altaffilmark{22}, 
C.~D. Wilson\altaffilmark{23}
}
\altaffiltext{1}{Department of Physics \& Astronomy, University of Wyoming, Laramie WY, USA; ddale@uwyo.edu}
\altaffiltext{2}{Cahill Center for Astronomy \& Astrophysics, California Institute of Technology, Pasadena CA, USA}
\altaffiltext{3}{Institut d'Astrophysique de Paris, Sorbonne Universit\'es, Paris, France}
\altaffiltext{4}{Spitzer Science Center, California Institute of Technology, Pasadena, CA, USA}
\altaffiltext{5}{Department of Astronomy, University of Maryland, College Park, MD, USA}
\altaffiltext{6}{Unidad de Astronom\'ia, Universidad de Antofagasta, Antofagasta, Chile}
\altaffiltext{7}{School of Physics \& Astronomy, Monash University, Victoria 3800, Australia}
\altaffiltext{8}{Department of Astronomy, University of Massachusetts, Amherst MA, USA}
\altaffiltext{9}{Sterrenkundig Observatorium, Universiteit Gent, Gent, Belgium}
\altaffiltext{10}{European Southern Observatory, Garching, Germany}
\altaffiltext{11}{Laboratoire AIM, CEA, Universit\'{e} Paris Diderot, IRFU/Service d'Astrophysique, Gif-sur-Yvette, France}
\altaffiltext{12}{Space Telescope Science Institute, Baltimore MD, USA}
\altaffiltext{13}{Research School of Astronomy \& Astrophysics, Australian National University, Canberra, Australia}
\altaffiltext{14}{Astronomy Department, University of Capetown, Rondebosch, South Africa}
\altaffiltext{15}{Max-Planck-Institut f\"ur Extraterrestrische Physik, Garching, Germany}
\altaffiltext{16}{Steward Observatory, University of Arizona, Tucson AZ, USA}
\altaffiltext{17}{INAF - Osservatorio Astrofisico di Arcetri, Firenze, Italy}
\altaffiltext{18}{Institute of Astronomy, University of Cambridge, Cambridge, UK}
\altaffiltext{19}{National Radio Astronomy Observatory, Charlottesville, VA, USA}
\altaffiltext{20}{Center for Astrophysics and Space Science, University of California, San Diego CA, USA}
\altaffiltext{21}{Department of Physics \& Astronomy, University of Toledo, Toledo, OH, USA}
\altaffiltext{22}{Instituto de Astrof\'{i}sica de Canarias, La Laguna, Spain}
\altaffiltext{23}{Department of Physics \& Astronomy, McMaster University, Hamilton, Ontario, Canada}

\begin {abstract}
We present an update to the ultraviolet-to-radio database of global broadband photometry for the 79 nearby galaxies that comprise the union of the KINGFISH (Key Insights on Nearby Galaxies: A Far-Infrared Survey with \Herschel) and SINGS (\Spitzer\ Infrared Nearby Galaxies Survey) samples.  The 34-band dataset presented here includes contributions from observational work carried out with a variety of facilities including \GALEX, \SDSS, \PS, \NOAO, \2MASS, \WISE, \Spitzer, \Herschel, \Planck, \JCMT, and the \VLA.  Improvements of note include recalibrations of previously-published SINGS \B\V\Rc\Ic\ and KINGFISH far-infrared/submillimeter photometry.  Similar to previous results in the literature, an excess of submillimeter emission above model predictions is seen primarily for low-metallicity dwarf/irregular galaxies.  This 33-band photometric dataset for the combined KINGFISH$+$SINGS sample serves as an important multi-wavelength reference for the variety of galaxies observed at low redshift.  A thorough analysis of the observed spectral energy distributions is carried out in a companion paper.
\end {abstract}

 

\section {Introduction}
\label{sec:intro}

Access to panchromatic broadband photometry for galaxies is crucial to fully understanding the characteristics of, and relative contributions to, galaxy spectra from the various processes related to interstellar attenuation, star formation, and the feeding of supermassive black holes \citep{silva98,dacunha08,boquien16}.  Though a fairly complete multi-wavelength dataset has been published for the SINGS (\Spitzer\ Infrared Nearby Galaxies Survey) sample of 75 nearby galaxies \citep{kennicutt03,dale05b,dale07}, subsequent far-infrared/submillimeter \Herschel\ broadband data were later published for the KINGFISH (Key Insights on Nearby Galaxies: A Far-Infrared Survey with \Herschel) sample of 61 nearby galaxies \citep{kennicutt11,dale12}, a sample for which 57 of the 61 targets are also SINGS targets.  The photometric datasets from the combined SINGS/KINGFISH surveys have served as important references for studies seeking to understand the diverse properties of galaxies in the Local Universe \citep{dacunha08,noll09,jonsson10,remyruyer15}, or to serve as redshift-zero comparison samples to higher redshift galaxies \citep{kartaltepe10,maiolino15,scoville16}.  

In this effort we present an update to the global (spatially-integrated) photometry for the 79 nearby galaxies that comprise the union of the KINGFISH and SINGS samples.  This update includes a recalibration of the \cite{dale12} KINGFISH far-infrared/submillimeter photometry, necessary since the calibration of the \Herschel\ photometers has undergone multiple revisions since those data were first published.  We also carry out a \PS-based recalibration of the \B\V\Rc\Ic\ fluxes previously published in \cite{dale07}.  A portion of those ground-based broadband optical data is suspect, since the data were originally taken in non-photometric conditions, and the ensuing attempts to calibrate the non-photometric frames were insufficient.  We also include new {\it ugriz} and 12\m\ photometry, respectively from the {\it Sloan Digital Sky Survey} (\SDSS) and the {\it Wide-Field Infrared Survey Explorer} (\WISE) mission.  In addition, the \Herschel\ PACS photometry for KINGFISH galaxy NGC~0584 was not included in \cite{dale12} since those imaging data were not yet taken.  Furthermore, we include new Herschel photometry for six SINGS galaxies from the \Herschel\ Very Nearby Galaxy Survey (PI C. Wilson) that are not in the KINGFISH sample (NGC~2403, M81=NGC~3031, M82=NGC~3034, NGC~4125, M51a=NGC~5194, and M51b=NGC~5195).  Finally, we complete the presentation of a multi-wavelength database by including previously published global photometry from ultraviolet (\GALEX), infrared/submillimeter (\2MASS, \Spitzer, SCUBA), and radio (\VLA) wavelengths.

Section~\ref{sec:sample} provides a brief overview of the galaxy sample, Section~\ref{sec:data} recaps the relevant observations and approaches to data processing, Section~\ref{sec:results} describes the salient results, and Section~\ref{sec:summary} provides the summary.  The companion paper (Hunt et al.\ 2017) explores the observed spectral energy distributions (SEDs) by utilizing and comparing fits using three popular fitting tools: GRASIL, MAGPHYS, and CIGALE \citep{silva98, dacunha08, noll09}.

\section {Galaxy Sample}
\label{sec:sample}

\begin{figure}
 \plotone{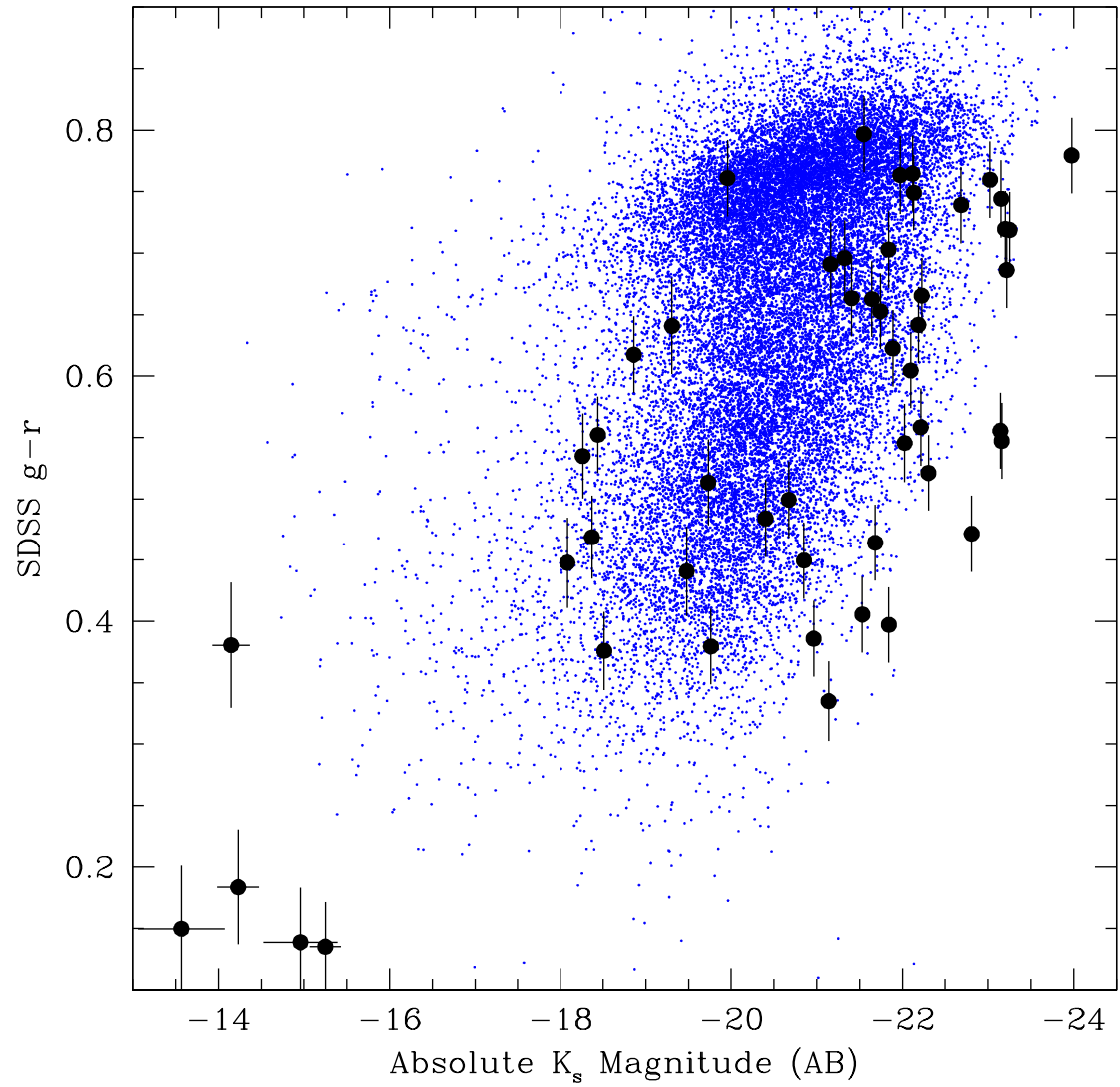}
 \caption{Comparison of global $g - r$ colors and absolute 2MASS \Ks\ magnitudes for the KINGFISH/SINGS (large circles) and \SDSS\ low redshift samples \citep[$10<d<150~{\rm Mpc}~h^{-1}$;][]{blanton05}.  The values are corrected for foreground Milky Way extinction.}
 \label{fig:greenvalley}
\end{figure}

Table~\ref{tab:sample} presents the full list of 79 galaxies that form the union of the SINGS and KINGFISH samples.  The sample was chosen to be a representative sampling of the Local Universe; the sample is not volume-limited and thus does not represent a statistical sampling of the Local Volume, but the sample is representative of the diversity of local galaxies.  These galaxies were selected to span a range of morphologies, colors, and luminosities \citep[e.g., Figures~5 and 3, respectively, of][]{kennicutt03,dale12}.   Figure~\ref{fig:greenvalley} demonstrates the sample's range of optical colors and near-infrared luminosities; a few galaxies reside in the red sequence near the top of the diagram but most of the sample spans the blue star-forming sequence.  The sample is comprised of 8\%, 11\%, 63\%, and 18\% early-type, lenticular, spiral, and irregular galaxies, respectively, based on the optical morphologies provided in the NASA/IPAC Extragalactic Database (NED).  This galaxy sample has no sources for which the optical luminosity is dominated by AGN emission, though one-third have signatures of Seyfert/LINER nuclei \citep{tajer05,dale06,moustakas10,grier11}.  There are only a few galaxies that are undoubtedly interacting with neighboring galaxies including NGC~5194 (with NGC~5195), NGC~1097 (with NGC~1097A), NGC~1316 (with NGC~1317), and NGC~3190 (with NGC~3187).  The distances reach out to $\sim30$~Mpc with a median value of $\sim10$~Mpc.  

\section {Observations and Data Processing}
\label{sec:data}

Much of the photometry presented here has already been described in \cite{dale07} and \cite{dale12}, so we focus the following discussion on important differences from what is presented in those publications.  The imaging bandpasses utilized here are listed in the heading of Table~\ref{tab:fluxes}. The central wavelengths and widths of the filters are computed via
\be
\bar{\lambda} \equiv {\int \lambda \mathcal{T}(\lambda) d\lambda \over \int \mathcal{T}(\lambda) d\lambda}
\label{eqn:filter_central}
\ee
\be
\Delta \equiv \int \mathcal{T}(\lambda) d\lambda
\label{eqn:filter_delta}
\ee
where $\mathcal{T}$ represents the filter transmittance normalized to peak at unity, based on the filter profiles compiled by \cite{noll09} for use in the CIGALE software package \citep{noll09}.  

\subsection{Ground-based Optical}
\label{sec:optical}

Some of the optical \B\V\Rc\Ic\ photometry from \cite{dale07} suffer from faulty calibration \citep{munozmateos09b}.  Sometimes insufficient numbers of standard star observations were taken; sometimes the standards were saturated; and some of the frames taken in non-photometric conditions were not successfully calibrated {\it a posteriori}.  A recalibration is carried out here via comparison of photometry on foreground stars in \PS\ (PS1) \gps\rps\ips\zps\ and in our 2007-era \B\V\Rc\Ic\ imaging.  The PS1 3$\pi$ Survey utilizes the 1.8~m telescope on Mount Haleakala to map the sky north of $\delta=-30\degr$ with multiple passes of 30--60~sec exposures in each of their five \SDSS-like filters \citep{schlafly12,magnier13}.

Care was taken in the comparison to only use bright (\rps$<$19~mag), unsaturated sources with point spread functions (PSFs) that agree with the seeing profiles of each image (i.e., background galaxies are excluded).  
The median number of foreground stars per galaxy utilized for this purpose was 15.  Aperture diameters for the foreground stellar photometry were typically 7\arcsec; increasing the aperture diameter by 50\% results in a $<$1\% difference in the calibration.  The PS1 fluxes were converted from their measured values at 1.2 airmasses to 0 airmasses \citep[Table~4 of][]{tonry12}, and the small (tens of millimag) corrections suggested by \cite{scolnic15} were incorporated (see their Table~3); the small calibration modifications of \cite{scolnic15} are based on a ``super calibration'' that combines flux measurements of secondary standards from several surveys including PS1 and \SDSS.  

We adopted the \cite{tonry12} quadratic filter transformations for stars between \PS\ \gps\rps\ips\zps\ and \B\V\Rc\Ic, though very similar results are obtained when using linear transformations between \SDSS\ and Johnson/Cousins filters \citep{jester05,lupton05,chonis08,jordi06,tonry12}.  The only significant outliers for any of these various stellar transformations are the \cite{chonis08} \V\ and \cite{lupton05} \Rc\ linear transformations, both of which yield calibrations that ultimately result in galaxy fluxes that are 25--30\% ($\sim 0.25-0.30$~mag) brighter compared to when using other published transformations.  The photometric calibrations are derived in practice from the error-weighted differences between the instrumental \B\V\Rc\Ic\ fluxes and the measured PS1 fluxes (transformed to \B\V\Rc\Ic) for the suite of suitable foreground stars identified for each galaxy.  The dispersions in these bootstrap calibrations range from 2\% to 15\% (with a median of 5\%) and contribute to the overall photometric uncertainty estimates; the photometric uncertainties for the PS1-recalibrated fluxes are the sums in quadrature of these dispersions along with the uncertainties in the published \gps\rps\ips\zps\ $\rightarrow$ \B\V\Rc\Ic\ transformations and the instrumental galaxy flux measurements.

The \PS\ survey does not encompass regions of the sky south of Galactic latitude $\delta = -30$\degr, and thus PS1 calibration is not possible for 18 SINGS/KINGFISH objects.  Table~\ref{tab:cal} indicates for which targets the \B\V\Rc\Ic\ photometry has been re-calibrated via PS1.  Additional broadband optical photometry is possible via other ground-based efforts.  \SDSS\ \u\g\r\i\z\ imaging (Data Release~12) is used to provide optical fluxes for 51 of the 79 SINGS/KINGFISH galaxies.  The union of the \SDSS\ \u\g\r\i\z\ and PS1-recalibrated \B\V\Rc\Ic\ samples comprises 63 galaxies, hence 80\% of the full sample.  The fraction of the sample for which we have reliable optical photometry approaches 100\% after inclusion of \B\V\Rc\Ic\ photometry from other global photometric datasets \citep[see Table~\ref{tab:cal} and][]{devaucouleurs91,munozmateos09b,tully09,cook14a}.

Other differences between our \B\V\Rc\Ic\ photometry and those appearing in \cite{dale07} include the use of SINGS Data Release~5 (DR5) images (DR2 imaging was used in the previous publication), more robust sky level determinations (i.e., a significantly larger number of sky pixels are now used---38\% more on average), and a fresh take on the editing of foreground stars and background galaxies (see \S~\ref{sec:skysub}).  In some instances the DR5 images are noticeably flatter than their DR2 counterparts (e.g., NGC~0628 \V\Rc\Ic).  Otherwise, the data processing procedures are essentially identical to those already described in \cite{dale07}.  

\subsection{\Herschel\ Infrared}
\label{sec:infrared}

Fluxes based on \Herschel\ PACS and SPIRE imaging are presented here for 67 of the 79 SINGS/KINGFISH galaxies.  The \Herschel\ PACS and SPIRE imaging observations are described in \cite{dale12} for the 61 KINGFISH galaxies, except for the PACS observations of NGC~0584 which were taken too late to 
appear in that publication.  Another minor difference from \cite{dale12} is that the PACS maps utilized here are deeper for five KINGFISH objects since we have now incorporated additional data from other observing programs: Holmberg~II, IC~2574, NGC~2798, NGC~4236, and NGC~4631.  Deeper observations allow for more robust measurements, e.g., the 70 and 100\m\ flux-to-uncertainty ratios for Holmberg~II are a factor of two larger than published in \cite{dale12} (see \S~\ref{sec:results}).  We also incorporate here new \Herschel\ imaging observations for six SINGS galaxies from the Herschel Very Nearby Galaxy Survey (VNGS; PI C. Wilson) that are not in the KINGFISH sample: NGC~2403, M81=NGC~3031, M82=NGC~3034, NGC~4125, M51a=NGC~5194, and M51b=NGC~5195.  The observing procedures for these six galaxies are described in \cite{bendo10}.

The \Herschel\ PACS and SPIRE imaging data for these $61+6=67$ KINGFISH + VNGS galaxies for this publication were processed from Level~0 to Level~1 using HIPE Version~11.1.0\footnote{Version~14.0.0 was used for the six VNGS galaxies.  This version provides better deglitched SPIRE maps and a slightly improved extended source calibration for PACS.}, and the Level~1 to Level~2 post-pipeline processing utilized Scanamorphos Version~24.0 \citep{roussel13}; the data published in \cite{dale12} were processed using HIPE Version~5.0 and Scanamorphos Version~12.5.  With this newer version of Scanamorphos, the PACS distortion flatfield is now properly incorporated.  In practice, this has decreased the noise levels in the PACS maps and 
slightly modified the PACS flux calibration ($\sim$1\%).
Moreover, the de-striping of PACS observations of large and diffuse fields is substantially improved, and the subtraction of the average drift on short timescales no longer introduces low-level noise.  These changes allow for more secure detections of diffuse emission and more robust estimates of sky levels.
 
One important factor involved in the SPIRE flux calibration is the SPIRE beam size, since the SPIRE images are converted to surface brightness units by dividing by the estimated beam areas.  The updated beam sizes used for this work at [250,350,500]$\mu$m are [469.35,831.27,1804.31]arcsec$^2$, representing percentage increases of [11.0,10.7,13.7]\% compared to the previous values of [423,751,1587]arcsec$^2$ used by \cite{dale12}.  These updated values are the recommended beam sizes in Version~3.0 (03 June 2016) of the SPIRE Handbook. 

The PACS fractional calibration uncertainties are of order $\epsilon_{\rm cal}/f_\nu \sim 5$\%, according to Version~2.5.1 (09~July~2013) of the PACS Observer's Manual.  Calibration uncertainties for SPIRE data are estimated at $\epsilon_{\rm cal}/f_\nu \sim 7$\%, also taken from Version~3.0 of the SPIRE Observer's Manual.  This level of uncertainty in the SPIRE calibration is a sum in quadrature of the uncertainties in the absolute and relative calibrations ($\sim$5.5\%) along with the uncertainties in the extended source calibration ($\sim$4\%).  


\subsection{\WISE\ 12\m}
\WISE\ 12\m\ imaging \citep{wright10} is utilized here to help bridge the gap in wavelength coverage in our broadband SEDs between the \Spitzer\ 8 and 24\m\ bandpasses.  The 12\m\ bandpass is an important tracer of the PAH complexes centered near (restframe) wavelengths of 11.3, 12.7, and 17\m\ \citep{smith07}.  At 12\m, the (single-frame) \WISE\ PSF full-width at half maximum (FWHM) is 6\farcs5 and the photometric calibration $\sim$7\%\footnote{Explanatory Supplement to the WISE All-Sky Data Release Products; 19 February 2015}.  A full suite of WISE W1, W2, W3, and W4 (3.4, 4.6, 12, and 22\m) ``total'' flux densities are provided in Table~\ref{tab:WISE} (see the Appendix).  We utilize native resolution imaging (not the drizzled Atlas release imaging) which provides superior resolution and minimizes the need for aperture corrections.

\subsection{\Planck\ 850\m}
There are SCUBA 850\m\ flux densities available for only 27 galaxies in the SINGS/KINGFISH sample and so we supplement these data with \Planck\ 850\m\ flux densities for 36 galaxies; the total number of 850\m\ detections for the combined \Planck/SCUBA SINGS/KINGFISH sample is 43.  We utilize cataloged \Planck\ 850\m\ flux densities derived from the APERFLUX technique, a technique that employs circular apertures plus concentric sky annuli.  Among the different flavors of \Planck\ flux density extractions, APERFLUX is the simplest method and relies on the fewest assumptions; the \Planck\ team recommends APERFLUX for wavelengths 850\m\ and shorter \citep{planck16a}.  Though some of the SCUBA 850\m\ data suffer from relatively small maps and spatial filtering issues, the lower-resolution \Planck\ data may be impacted by foreground/background sources.

\subsection{Sky Estimation and Elimination of Spatial Interlopers} 
\label{sec:skysub}

The ``sky'' in the direction of any galaxy is a superposition of emission from faint foreground stars and background galaxies, interstellar emission, and in the case of ground-based observations, the Earth's atmosphere.  To determine the sky level for each image, a set of sky apertures has been defined (by eye) that collectively circumscribe the galaxy, projected on the sky close enough to the galaxy to measure the ``local'' sky but far enough away to avoid any galaxy emission \citep[the process is unchanged from][see their Figure~1]{dale12}.  

Prior to estimating the sky levels and executing aperture photometry, any emission from obvious foreground stars or neighboring/background galaxies is identified and removed from the areas covered by each galaxy's aperture and collection of sky apertures.  The identification is assisted by ancillary data at shorter wavelengths and higher spatial resolution (e.g., {\it Spitzer}/IRAC 3.6 and 8.0\m, {\it HST} optical, and ground-based \hal\ imaging).  The removal is accomplished via IRAF/{\tt IMEDIT} by replacing the values of contaminated pixels via a 2D surface fit to to a nearby sky annulus of width 2~pixels, with noise added that matches the noise statistics of the sky annulus.  These annuli only sample the local sky around each spatial interloper and thus do not capture the full sky variations across the image.  Fortunately, the spatial interlopers are much smaller than our target galaxies, a fact that limits the impact of any shortcomings in the contaminant removal procedure. 

The total sky area, derived from the sum of the areas from all sky apertures, is typically significantly greater than that covered by the galaxy aperture itself, thereby limiting the contribution of uncertainty in the sky level to the overall error budget.  The mean sky level per pixel is computed from the collection of these sky apertures, the value is scaled to the number of pixels in the galaxy aperture, and the result is subtracted off from the overall galaxy aperture counts.  

\subsection{Aperture Photometry}
\label{sec:ap_phot}

The elliptical apertures used for global photometry are listed in Table~\ref{tab:sample}, and the same aperture is used to extract the flux at each wavelength.  The apertures were chosen to encompass essentially all of the detectable emission at every wavelength \citep[see also][]{dale05b,dale07,dale12}.  The average ratio of aperture major axis length $2a$ to the de Vaucouleurs $D_{25}$ optical major axis is 1.45 (with a 1$\sigma$ dispersion in this ratio of 0.45).

At the longest wavelengths where the imaging resolution is typically coarsest, a small portion of the galaxy emission may appear beyond the chosen apertures.  Thus, for the \Spitzer\ and \Herschel\ photometry we utilize the aperture corrections described in \cite{dale07} and \cite{dale12}.  No aperture corrections were applied to the WISE photometry as they are negligible for the large apertures used here on native resolution imaging.

The uncertainties in the integrated photometry $\epsilon_{\rm total}$ are computed as a combination in quadrature of the calibration uncertainty $\epsilon_{\rm cal}$ and the measurement uncertainty $\epsilon_{\rm sky}$ based on the measured sky fluctuations and the areas covered by the galaxy and the sum of the sky apertures, i.e., 
\be
\epsilon_{\rm total} = \sqrt{\epsilon_{\rm cal}^2 + \epsilon_{\rm sky}^2}
\ee
with
\be
\epsilon_{\rm sky} ~ = \sigma_{\rm sky} \Omega_{\rm pix} ~ \sqrt{N_{\rm pix}+{N_{\rm pix}}^2/N_{\rm sky}}
\ee
where $\sigma_{\rm sky}$ is the standard deviation of the sky surface brightness fluctuations, $\Omega_{\rm pix}$ is the solid angle subtended per pixel, and $N_{\rm pix}$ and $N_{\rm sky}$ are the number of pixels in the galaxy and (the sum of) the sky apertures, respectively.  For the few sources undetected by \Spitzer, \Herschel, or \WISE\ imaging, 5$\sigma$ upper limits are derived assuming a galaxy spans all $N_{\rm pix}$ pixels in the aperture,
\be
f_\nu(5\sigma~{\rm upper~limit}) ~ = ~ 5 ~ \epsilon_{\rm sky}.
\ee
Based on our visual scrutiny of the imaging datasets from a given telescope, any images redward of a non-detection are also considered to yield non-detections.

The galaxy apertures, the sky apertures, and the foreground stellar masks are provided with the electronic version of the journal article.

\section {Results}
\label{sec:results}

\subsection {Flux Densities} 

Table~\ref{tab:fluxes} presents the spatially-integrated flux densities for all 79 SINGS+KINGFISH galaxies for 30 photometric bands.  In Table~\ref{tab:upperlimits} we also supply global aperture photometry for the few cases where upper limits are provided in Table~\ref{tab:fluxes}.  The tabulated flux densities include aperture corrections (\S~\ref{sec:ap_phot}) and are {\it not} corrected for Galactic extinction.  No color corrections have been applied to the data in Table~\ref{tab:fluxes}.  Some of the fluxes presented here remain unchanged from the values published elsewhere, for example \2MASS\ \J\H\Ks\, \Spitzer\ IRAC and MIPS, SCUBA\ 850\m, and VLA 20~cm photometry.  However, if any differences exist between values published in multiple publications, precedence is given to the more recent published values, e.g., \2MASS\ and \Spitzer\ photometry from the Local Volume Legacy publication of \cite{dale09b} is given priority over the \2MASS\ and \Spitzer\ photometry appearing in \cite{dale07}.

\begin{figure}
 \plotone{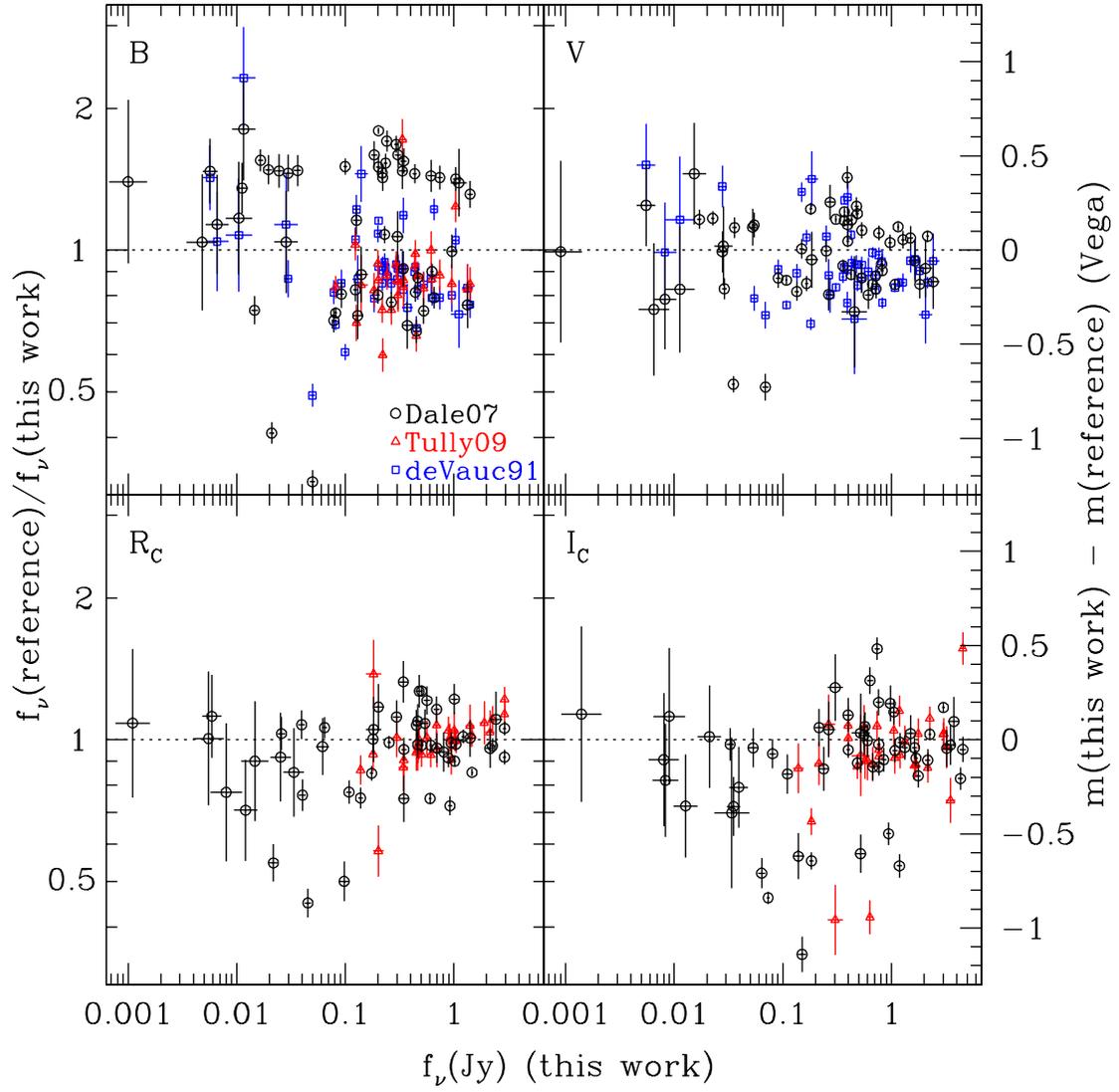}
 \caption{Comparison of global {\it BVR$_{\rm C}$I$_{\rm C}$} galaxy photometry from the literature with those measured here which are calibrated based on Pan-STARRS1 {\it g$_{\rm P1}$r$_{\rm P1}$i$_{\rm P1}$z$_{\rm P1}$} photometry on field stars (see \S~\ref{sec:optical}).}
 \label{fig:drpp}
\end{figure}

Figure~\ref{fig:drpp} provides a comparison of our updated optical fluxes with those presented in previous publications.  The updated \V\Rc\Ic\ optical fluxes are generally in agreement, on average, with previously-published values, and the 1$\sigma$ scatters in the differences for these filters are $\sim$0.2--0.3~mag.  The differences with published \B\ filter fluxes, however, show a locus of points indicating the literature data are typically 0.2~mag fainter than what we obtain after calibrating via \PS.  There is also a second grouping of \B\ data points, comprising nearly one-third of the total sample, that indicates the \cite{dale07} data are $\sim$0.4~mag {\it brighter}; these may be cases where the \B\ standard star calibration images for \cite{dale07} were either saturated or obtained in non-photometric conditions, both of which would lead to artificially faint standard star counts/sec and thus artificially bright galaxy fluxes.

\begin{figure}
 \plotone{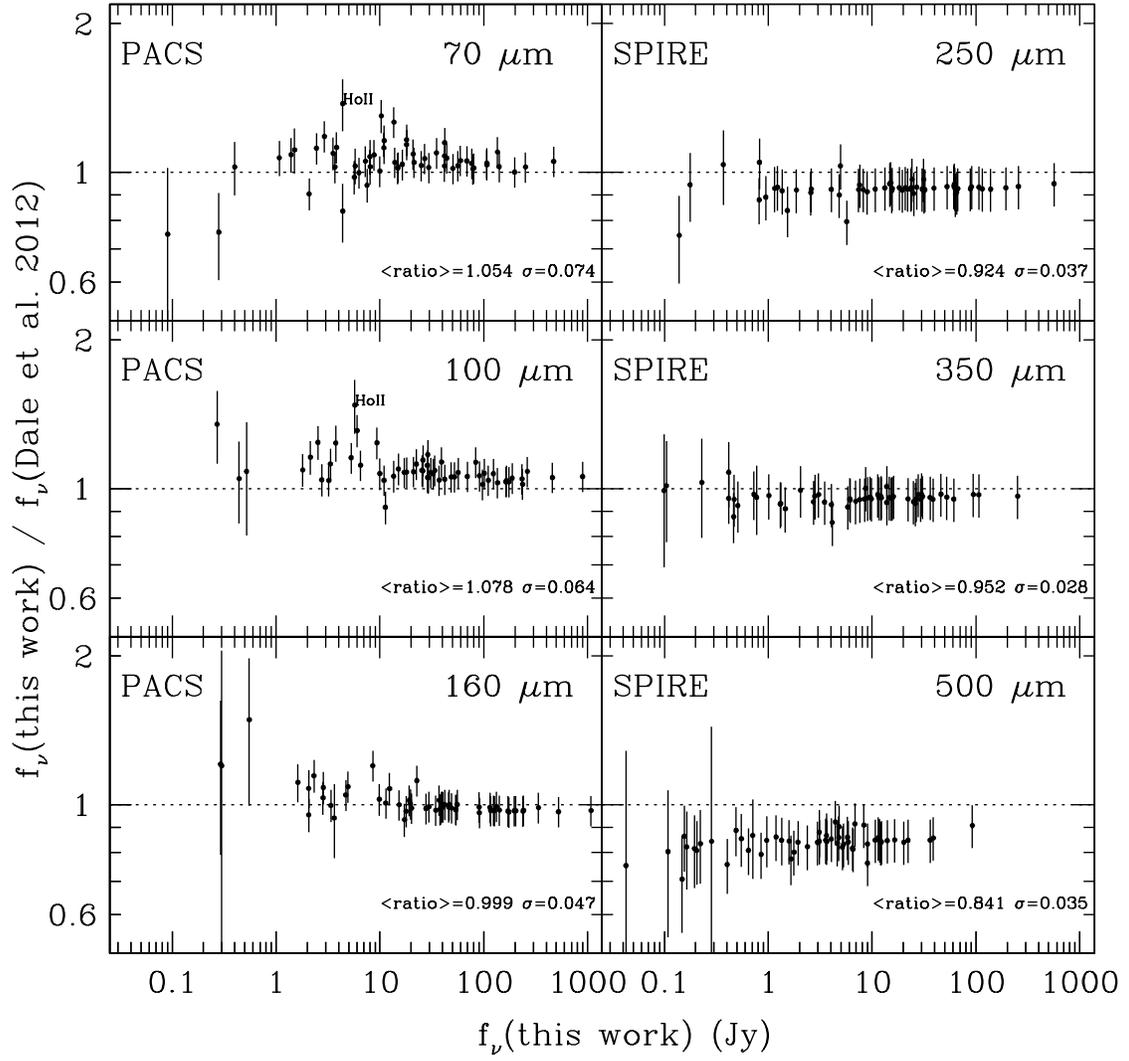}
 \caption{Comparison of updated global \Herschel\ photometry with those presented in \cite{dale12}.  The error bars are relatively constant since they are primarily dominated by systematics for the brighter sources.}
 \label{fig:pacsspire}
\end{figure}

Figure~\ref{fig:pacsspire} provides a similar comparison for the \Herschel\ far-infrared/submillimeter photometry.  The updated SPIRE beam sizes are larger than those used in \cite{dale12} by [11,11,14]\% at [250,350,500]$\mu$m, which naturally leads to fainter SPIRE fluxes.  This decrease is evident in Figure~\ref{fig:pacsspire}, where the average SPIRE flux is $\sim$[8,5,16]\% fainter at [250,350,500]$\mu$m than what appeared in \cite{dale12}.  For the five galaxies where we incorporated additional PACS data from other observing programs (Holmberg~II, IC~2574, NGC~2798, NGC~4236, and NGC~4631; see \S~\ref{sec:data}), the resulting PACS maps are deeper and thus allow for some diffuse flux to be additionally detected.  For faint Holmberg~II it makes an appreciable difference: the 70 and 100\m\ global fluxes are now 40-50\% larger.  

\subsection {The Spectral Energy Distributions} 
\label{sec:fits}

\begin{figure}
 \plotone{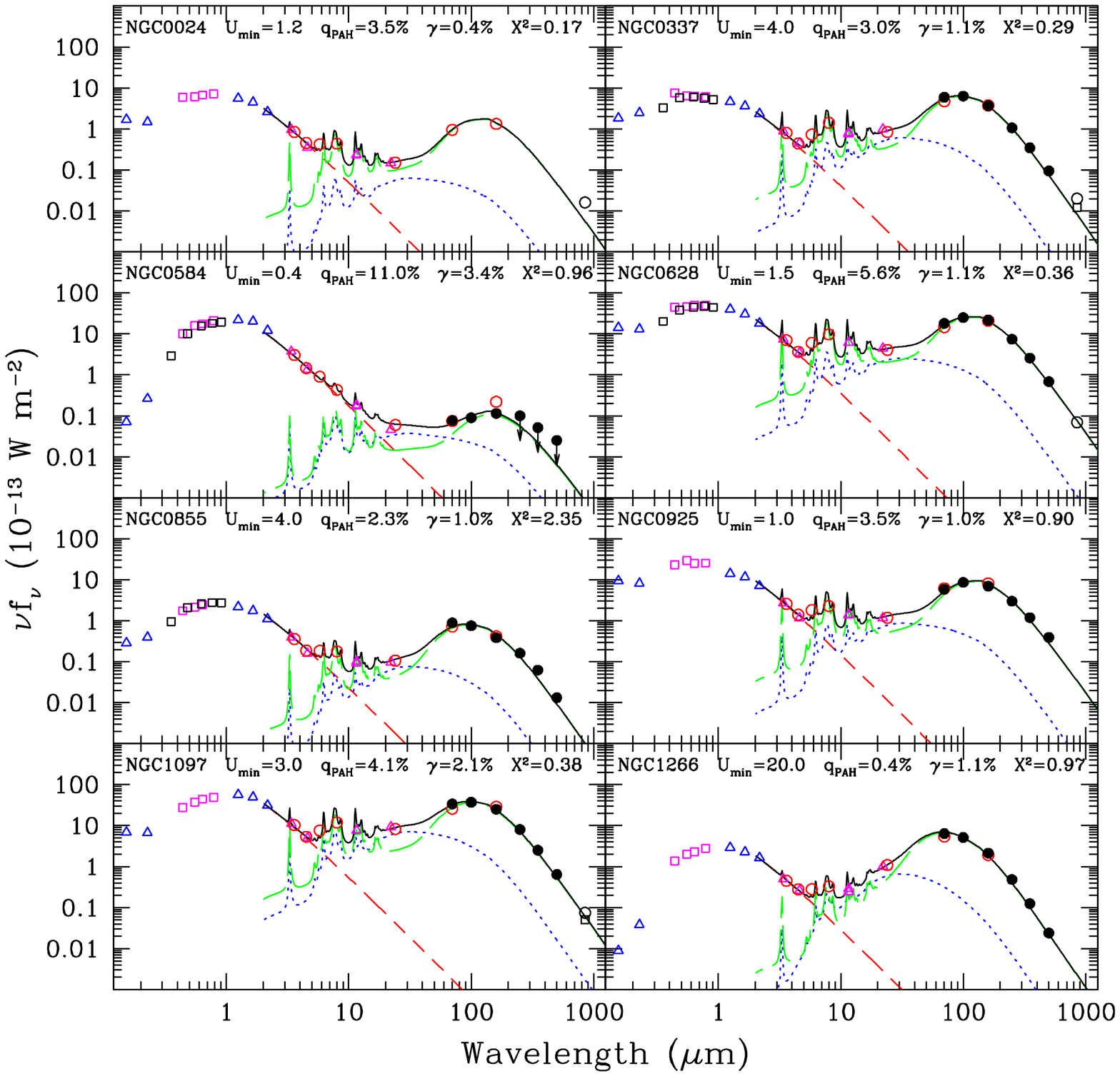}
 \caption{Globally-integrated infrared/sub-millimeter spectral energy distributions for all the galaxies in the KINGFISH/SINGS sample, sorted by Right Ascension.  The following symbols are utilized: filled circles (\Herschel), triangles (\GALEX, \2MASS, and \WISE), open circles (\Spitzer\ and \Planck), squares (\B\V\Rc\Ic\ and \u\g\r\i\z\ and SCUBA).  Arrows indicate 5$\sigma$ upper limits (and lower limits in the case of NGC~3034).  The solid curve is the sum of a 5000~K stellar blackbody (short dashed) along with models of dust emission from PDRs (dotted; $U>U_{\rm min}$) and the diffuse interstellar medium (long dashed; $U=U_{\rm min}$).  The fitted parameters from these \cite{draineli07} 3.6--500\m\ model fits are listed within each panel along with the reduced $\chi^2$ (see \S~\ref{sec:fits} for details).  While the plotted data are corrected for Galactic extinction, the fluxes tabulated in Table~\ref{tab:fluxes} are not corrected.  The uncertainties are smaller than the symbols plotted.  There are no SED fits in cases where the far-infrared photometry provides only limits.}
 \label{fig:seds}
\end{figure}

Figure~\ref{fig:seds} shows the observed infrared/sub-millimeter spectral energy distributions for the KINGFISH sample.  Included in each panel, when available, are the \GALEX\ far- and near-ultraviolet, \B\V\Rc\Ic\ and \u\g\r\i\z\ optical, \2MASS\ \J\H\Ks\ near-infrared, \Spitzer\ 3.6, 4.5, 5.8, 8.0, 24, 70, and 160\m, \WISE\ 12\m, \Herschel\ 70, 100, 160, 250, 350, and 500\m, and \Planck\ and SCUBA\ 850\m\ fluxes. 

The broadband spectral energy distributions displayed in Figure~\ref{fig:seds} are fitted with the models of \cite{draineli07} over the wavelength range 3.6--500\m, models based on mixtures of amorphous silicate and graphitic dust grains that effectively reproduce the average Milky Way extinction curve and are consistent with observations of PAH features and the variety of infrared continua in local galaxies.  A total of four free parameters are utilized in the fits: the fraction $q_{\rm PAH}$ of the dust mass residing in polycyclic aromatic hydrocarbons (PAHs), the intensity $U_{\rm min}$ of the interstellar radiation field that heats the general diffuse interstellar medium, and the fraction $\gamma$ of the dust mass heated by more intense starlight distributions such as those arising from photodissociation regions (PDRs) in star-forming regions, the ratio of the stellar mass to the dust mass ($M_{\rm dust}$ is determined by the normalization of the SED model with the observed photometric data).  The parameter $q_{\rm PAH}$ ranges between 0\% and 12\%, and $U_{\rm min}$ can have values between 0.01 and 30.  As was done for \cite{draine07} and \cite{dale12}, we minimize the number of free parameters by fixing the maximum value of the interstellar radiation field ($U_{\rm max}=10^6$) as well as the power-law exponent that governs the distribution of starlight intensities heating the dust ($\alpha=2$).  More details of these models may be found in \cite{draineli07} and \cite{draine07}.  

The new SED fits here can be compared to what was obtained previously for the KINGFISH sample in \cite{dale12} with the outdated SPIRE photometric calibration.  Figure~\ref{fig:DL07} provides such a comparison for dust mass, PAH fraction, and the properties of the radiation field that is heating the dust.  The two systematic differences, in the inferred dust mass and the diffuse radiation field intensity $U_{\rm min}$, are the result of the changes in the PACS and SPIRE calibrations---brighter at 70 and 100\m\ and fainter at 250, 350, and 500\m\ result in warmer interstellar dust and smaller overall dust masses.  We note that, while recent results indicate a necessary change in the DL07 dust opacities \citep{dalcanton15,planck16b}, we are consistently using the same (original) dust models from \cite{draineli07} in the Figure~\ref{fig:DL07} comparisons.  The two outliers in $q_{\rm PAH}$ are the result of the inclusion of new PACS photometry for NGC~0584 and an improved sky estimate for NGC1291's PACS 70\m\ map, which led to a factor of $\sim$2 smaller flux that is now in much better agreement with the \Spitzer\ 70\m\ flux.  

As was mentioned in \cite{dale12}, the faintest targets pose the most challenges to SED fitting.  Spatial variations in the foreground cirrus amplify the relative uncertainty in the extracted fluxes for such targets.  Even with our updated data processing and analysis, this situation remains essentially unchanged for galaxies like NGC~0584, DDO~053, M81~dwarf~B, and Holmberg~I.  However, the addition of \WISE\ fluxes to the SED fitting incrementally improves our confidence in the SED fits.  A more detailed analysis of theoretical fits to these SEDs is presented in the companion paper (Hunt et al.\ 2017) and in Draine et al. (2017).

\subsection {Sub-millimeter Excess}
\label{sec:submm}

In \cite{dale12} we found an excess of submillimeter emission for several galaxies in the KINGFISH sample, where the observed 500\m\ emission was measured to lie significantly above the model predictions based on \cite{draineli07} fits to the observed 3.6--500\m\ SEDs.  The submillimeter excess $\xi(500\mu{\rm m})$ was quantitatively defined in \cite{dale12} as
\be
\xi(500\mu{\rm m}) = {f_\nu(500\mu{\rm m})_{\rm observed} - f_\nu(500\mu{\rm m})_{\rm model} \over f_\nu(500\mu{\rm m})_{\rm model}}.
\label{eq:excess}
\ee
This excess was primarily found in low-metallicity galaxies: nine of the ten dwarf/irregular/Magellanic galaxies in the KINGFISH sample with 500\m\ detections exhibited $\xi(500\mu{\rm m})>0.60$.  This result echoed similar claims of submillimeter excess in studies of M~33, the Magellanic Clouds, and other low-metallicity star-forming galaxies \citep{galliano05,planck11,galametz11,gordon14,izotov14,hermelo16}.  With the updated calibrations for our present dataset, including SPIRE calibration changes that result in 500\m\ fluxes lower by an average of 16\% (Figure~\ref{fig:pacsspire}), we still see submillimeter excesses in the sample but at lower significance.  The average excess for dwarf/irregular/Magellanic galaxies in \cite{dale12} was $\left< \xi(500\mu{\rm m})\right>_{\rm T=Im,I0,Sm} \sim 0.70$, whereas in the current work the average value is $\left< \xi(500\mu{\rm m})\right>_{\rm T=Im,I0,Sm} \sim 0.53$ for the 11 SINGS/KINGFISH dwarf/irregular/Magellanic galaxies with secure 500\m\ detections.  Note that an alternative signal-to-noise-like definition of the submillimeter excess may be defined by normalizing via the model and observational uncertainties, namely 
\be
S/N(500\mu{\rm m}~{\rm excess}) = {f_\nu(500\mu{\rm m})_{\rm observed} - f_\nu(500\mu{\rm m})_{\rm model} \over \sqrt{\epsilon_{\rm obs}^2 + \epsilon_{\rm mod}^2}}.
\ee
In this case the average value is $\langle S/N(500\mu{\rm m}~{\rm excess}) \rangle \sim 4.3$ for $\epsilon_{\rm mod}=0$ and $\sim 3.3$ for $\epsilon_{\rm mod}=0.1 f_\nu(500\mu{\rm m})_{\rm model}$ for the 11 SINGS/KINGFISH dwarf galaxies.  We caution that our results may be biased since we restrict our analysis to securely-detected sources.

Though the lefthand panel of Figure~\ref{fig:excess} shows no dependence for submillimeter excess on far-infrared color, there is a clear trend in the righthand panel, with higher submillimeter excesses for lower gas-phase metallicity.  This persistency of a submillimeter excess in primarily low-metallicity galaxies conflicts with the analysis of \cite{kirkpatrick13}, who found no significant submillimeter excesses for a subset of 20 galaxies from the KINGFISH sample with robust (and updated) far-infrared photometry that span a range in metallicity.  However, there are three main differences between the analysis carried out here and by \cite{kirkpatrick13}.  First, \cite{kirkpatrick13} only fit data spanning 24--350\m.  Our SED fitting employs the full infrared continuum over 3.6--500\m\ and therefore our approach must balance contributions from stellar and PAH emission in addition to that from larger dust grains.  Second, \cite{kirkpatrick13} employ the superposition of two modified blackbodies whereas our fits utilize a more sophisticated dust model \citep{draineli07}; we have four free parameters in our SED fitting whereas \cite{kirkpatrick13} have only three free parameters: the temperature and emissivity of the cold dust component and the ratio of the amplitudes of the two modified blackbodies.  Third, while we normalize our submillimeter excesses by the predicted model value at 500\m, \cite{kirkpatrick13} normalize via the {\it observed} 500\m\ value.  Normalizing by the observed flux is an approach that naturally depresses the excess measure.  For example, the average submillimeter excess for the 11 SINGS/KINGFISH dwarf/irregular/Magellanic galaxies with 500\m\ detections is 0.33 if the normalization is via the observed flux, a factor of 1.6 times smaller than when normalizing by the predicted model flux.

\section {Summary}
\label{sec:summary}

We present an update to the full panchromatic photometric database for the 79 galaxies in the combined SINGS/KINGFISH sample of nearby galaxies, for ultraviolet through radio wavelengths.  Updates include incorporating recent improvements in the calibration of the \Herschel\ photometers and a re-calibration of \B\V\Rc\Ic\ photometry utilizing broadband data from the \PS\ survey.  On average, the updated \Herschel\ fluxes differ by [$+$5,$+$8,0,$-$8,$-$5,$-$16]\% at [70,100,160,250,350,500]$\mu$m compared to what was published in the original KINGFISH photometry paper of \cite{dale12}.  The average updated fluxes for the \V\Rc\Ic\ filters are essentially unchanged from the SINGS photometry presented in \cite{dale07}, but with a scatter of 0.2--0.3~mag.  The updated collection of \B\ fluxes are about $\sim$0.2~mag brighter than what appears in the literature. Finally, theoretical \cite{draineli07} SED models are fit to each galaxy's 3.6--500\m\ dataset.  Two of the fitted parameters show small but systematic differences with the results published previously: the total dust masses are about 3\% smaller and the typical value of the radiation field that is heating the diffuse ISM is about 16\% smaller. Both of these results naturally arise from the changes in the calibrations of the \Herschel\ imagers since 2012.  We confirm our previous finding of an excess of submillimeter emission (500\m) in primarily low-metallicity dwarf/irregular galaxies, but at a smaller amplitude due to the updated calibration of the \Herschel\ SPIRE beams.  A full exploration of this panchromatic dataset is carried out in a companion paper by Hunt et al.\ (2017).
 
\clearpage

We thank the referee for excellent suggestions that improved this work.
{\em Herschel} is an ESA space observatory with science instruments provided by European-led Principal Investigator consortia and with important participation from NASA.  This work is based on observations made with the {\it Spitzer Space Telescope} and utilizes the NASA/IPAC Infrared Science Archive, both operated by JPL/Caltech under a contract with NASA.  We gratefully acknowledge NASA's support for construction, operation, and science analysis for the GALEX mission, developed in cooperation with the Centre National d'Etudes Spatiales of France and the Korean Ministry of Science and Technology.  Funding for the Sloan Digital Sky Survey and SDSS-II has been provided by the Alfred P. Sloan Foundation, the Participating Institutions, the NSF, the U.S. Department of Energy, NASA, the Japanese Monbukagakusho, the Max Planck Society, and the Higher Education Funding Council for England.
This publication makes use of data products from the Wide-field Infrared Survey Explorer, which is a joint project of the University of California, Los Angeles, and the JPL/Caltech, funded by NASA.  The Pan-STARRS1 Surveys have been made possible through contributions of the Institute for Astronomy, the University of Hawaii, the Pan-STARRS Project Office, the Max Planck Society and its participating institutes, the Max Planck Institute for Astronomy, Heidelberg and the Max Planck Institute for Extraterrestrial Physics, Garching, The Johns Hopkins University, Durham University, the University of Edinburgh, Queen's University Belfast, the Harvard-Smithsonian Center for Astrophysics, the Las Cumbres Observatory Global Telescope Network Incorporated, the National Central University of Taiwan, the Space Telescope Science Institute, the NSF, the University of Maryland, and Eotvos Lorand University and the Los Alamos National Laboratory.

\appendix{Appendix: WISE Fluxes}

Table~\ref{tab:fluxes} presents WISE W3 (12\m) flux densities based on the same apertures used for the other flux densities presented in Table~\ref{tab:fluxes}.  We additionally present here ``total'' flux densities for all four WISE bands using zeropoints of [309.666,170.623,29.043,7.875]~Jy for [W1,W2,W3,W4] at central wavelengths of [3.4,4.6,12,22]$\mu$m\footnote{\cite{brown14} suggest W4 has an effective central wavelength of 22.8\m}.  The process for extracting ``total'' flux densities follows that described in \citep{jarrett13}, whereby azimuthally-averaged elliptical surface brightness profiles are extrapolated to three disk scale lengths beyond the 1$\sigma$ (sky rms) isophotal radii.  For WISE, the 1$\sigma$ isophotes are at surface brightness levels of approximately [23.0,21.8,18.1,15.8]~mag~arcsec$^{-2}$ (Vega) for [W1,W2,W3,W4] \citep{jarrett13}.  This particular WISE database provides an excellent opportunity to check our \Spitzer- and aperture-based global fluxes at 3.6, 4.5, and 24\m.  Figure~\ref{fig:wise} presents such a comparison.  The agreement between \Spitzer\ and \WISE\ flux densities is close to expectations.  The dotted lines in the top two panels indicate the expected flux density ratios after convolving the filter bandpasses with a 5000~K stellar blackbody; the observed ratios match the expected ratios to within the observed 1$\sigma$ scatters.  There is perhaps a weak trend for the ratio of \Spitzer\ 24\m\ and \WISE\ W3, with the ratio decreasing with increasing brightness.  


\begin{deluxetable}{lllccrrrrr}
\tablenum{1}

\def\p{$\pm$}
\tabletypesize{\scriptsize}
\tablecaption{Galaxy Sample}
\tablewidth{0pc}
\tablehead{
\colhead{Galaxy} &
\colhead{Alternative} &
\colhead{Optical} &
\colhead{$E(B-V)$} &
\colhead{$\alpha_0$~\&~$\delta_0$} &
\colhead{2$a$} &
\colhead{2$b$} &
\colhead{PA} &
\colhead{$D$} &
\colhead{TIR} 
\\
\colhead{} &
\colhead{Name} &
\colhead{Morph.} &
\colhead{(mag.)} &
\colhead{(J2000)} &
\colhead{($\arcsec$)} &
\colhead{($\arcsec$)} &
\colhead{($\degr$)} &
\colhead{(Mpc)} &
\colhead{($L_\odot$)} 
}
\startdata
NGC0024   &UGCA002    &SAc   &0.017&000955.9$-$245755& 301& 216& 45& 8.20&    8.8\\
NGC0337   &NGC0337    &SBd   &0.096&005950.7$-$073444& 253& 194&140&19.30&   10.1\\
NGC0584   &NGC0584    &E4    &0.036&013120.6$-$065205& 326& 278& 60&20.80&    8.7\\
NGC0628   &Messier074 &SAc   &0.060&013642.4$+$154711& 879& 808& 90& 7.20&    9.9\\
NGC0855   &UGC01718   &E     &0.061&021403.7$+$275237& 259& 169& 60& 9.73&    8.6\\
NGC0925   &UGC01913   &SABd  &0.065&022713.6$+$333504& 735& 486&105& 9.12&    9.7\\
NGC1097   &UGCA041    &SBb   &0.023&024618.0$-$301642& 758& 612&130&14.20&   10.6\\
NGC1266   &NGC1266    &SB0   &0.085&031600.6$-$022541& 234& 232&  0&30.60&   10.4\\
\enddata
\tablecomments{\footnotesize Foreground extinctions derive from \cite{schlafly11}.  $2a$ and $2b$ are the lengths of the major and minor axes used in the elliptical aperture photometry described herein; the position angle of the aperture's major axis is measured east of north.  The total infrared luminosity listed in the last column is derived from Equation~5 of \cite{dale14} and the \Spitzer\ 8, 24, 70, and 160\m\ fluxes in Table~\ref{tab:fluxes} and assumes the distances provided in this table.  \WISE\ 22\m\ and \Herschel\ PACS 70 and 160\m\ fluxes are used for NGC~3034 since its MIPS data suffer from saturation.}
\label{tab:sample}
\end{deluxetable}

\clearpage
\begin{deluxetable}{lrrrrrrrrrr}
\tablenum{2}
\def\a{\tablenotemark{a}}
\def\p{$\pm$}
\floattable
\rotate
\tabletypesize{\tiny}
\tablecaption{Global Flux Densities in Janskys Corrected for Neither Galactic Nor Intrinsic Extinction}
\tablewidth{0pc}
\tablehead{
\colhead{} &
\colhead{\GALEX} &
\colhead{\GALEX} &
\colhead{} &
\colhead{} &
\colhead{} &
\colhead{} &
\colhead{\SDSS} &
\colhead{\SDSS} &
\colhead{\SDSS} &
\colhead{\SDSS} 
\\
\colhead{Filter} &
\colhead{FUV} &
\colhead{NUV} &
\colhead{\B} &
\colhead{\V} &
\colhead{\Rc} &
\colhead{\Ic} &
\colhead{\u} &
\colhead{\g} &
\colhead{\r} &
\colhead{\i} 
\\
\colhead{$\bar{\lambda}$($\mu$m)} &
\colhead{0.154} &
\colhead{0.231} &
\colhead{0.441} &
\colhead{0.551} &
\colhead{0.659} &
\colhead{0.806} &
\colhead{0.356} &
\colhead{0.482} &
\colhead{0.626} &
\colhead{0.767} 
\\
\colhead{$\Delta \lambda$($\mu$m)} &
\colhead{0.025} &
\colhead{0.073} &
\colhead{0.096} &
\colhead{0.089} &
\colhead{0.159} &
\colhead{0.150} &
\colhead{0.056} &
\colhead{0.127} &
\colhead{0.133} &
\colhead{0.135} 
\\
\colhead{$A_\lambda/A_V$} &
\colhead{2.586} &
\colhead{2.994} &
\colhead{1.310} &
\colhead{1.0} &
\colhead{0.788} &
\colhead{0.577} &
\colhead{1.642} &
\colhead{1.219} &
\colhead{0.849} &
\colhead{0.641} 
}
\startdata
NGC0024&772\p121E$-$3&981\p149E$-$3&817\p021E$-$2&108\p002E$-$1&138\p007E$-$1&183\p007E$-$1&\nodata      &\nodata      &\nodata      &\nodata      \\
NGC0337&463\p064E$-$3&819\p113E$-$3&786\p046E$-$2&908\p044E$-$2&108\p006E$-$1&139\p015E$-$1&249\p006E$-$2&664\p014E$-$2&100\p002E$-$1&121\p002E$-$1\\
NGC0584&274\p039E$-$4&146\p020E$-$3&130\p014E$-$1&264\p020E$-$1&346\p039E$-$1&519\p048E$-$1&292\p007E$-$2&143\p002E$-$1&296\p006E$-$1&433\p008E$-$1\\
NGC0628&469\p073E$-$2&591\p091E$-$2&528\p033E$-$1&721\p033E$-$1&921\p042E$-$1&118\p006E$+$0&179\p003E$-$1&492\p009E$-$1&804\p016E$-$1&108\p002E$+$0\\
NGC0855&920\p144E$-$4&178\p027E$-$3&204\p020E$-$2&335\p033E$-$2&457\p012E$-$2&\nodata      &840\p029E$-$3&268\p006E$-$2&468\p010E$-$2&633\p013E$-$2\\
NGC0925&293\p040E$-$2&357\p049E$-$2&265\p013E$-$1&456\p123E$-$1&465\p039E$-$1&601\p049E$-$1&\nodata      &\nodata      &\nodata      &\nodata      \\
NGC1097&298\p042E$-$2&418\p058E$-$2&373\p013E$-$1&641\p023E$-$1&898\p033E$-$1&124\p004E$+$0&\nodata      &\nodata      &\nodata      &\nodata      \\
NGC1266&024\p003E$-$4&141\p019E$-$4&146\p010E$-$2&289\p015E$-$2&404\p031E$-$2&638\p047E$-$2&\nodata      &\nodata      &\nodata      &\nodata      \\
\enddata     
\tablecomments{\footnotesize The compact table entry format TUV$\pm$WXYEZ implies (T.UV$\pm$W.XY)$\times10^{\rm Z}$ in Jy.  Corrections for neither Galactic nor intrinsic extinction has been applied (see \S~\ref{sec:data}).  The uncertainties include both statistical and systematic effects.  5$\sigma$ upper limits are provided for non-detections.  No color corrections have been applied.  The filter central wavelengths and widths are computed via Equations~\ref{eqn:filter_central} and \ref{eqn:filter_delta}.}
\tablenotetext{a}{\footnotesize The \Spitzer\ MIPS data for NGC~3034 suffer from saturation and thus the tabulated values are lower limits.}
\label{tab:fluxes}
\end{deluxetable}

\clearpage
\begin{deluxetable}{lrrrrrrrrrr}
\tablenum{2}
\def\a{\tablenotemark{a}}
\def\p{$\pm$}
\floattable
\rotate
\tabletypesize{\tiny} 
\tablecaption{Global Flux Densities in Janskys Corrected for Neither Galactic Nor Intrinsic Extinction}
\tablewidth{0pc}
\tablehead{
\colhead{} &
\colhead{\SDSS} &
\colhead{\2MASS} &
\colhead{\2MASS} &
\colhead{\2MASS} &
\colhead{\Spitzer} &
\colhead{\Spitzer} &
\colhead{\Spitzer} &
\colhead{\Spitzer} &
\colhead{WISE} &
\colhead{\Spitzer} 
\\
\colhead{Filter} &
\colhead{\z} &
\colhead{\J} &
\colhead{\H} &
\colhead{\Ks} &
\colhead{IRAC} &
\colhead{IRAC} &
\colhead{IRAC} &
\colhead{IRAC} &
\colhead{W3} &
\colhead{MIPS} 
\\
\colhead{$\bar{\lambda}$($\mu$m)} &
\colhead{0.910} &
\colhead{1.24} &
\colhead{1.65} &
\colhead{2.17} &
\colhead{3.56} &
\colhead{4.51} &
\colhead{5.76} &
\colhead{7.96} &
\colhead{12.8} &
\colhead{23.8} 
\\
\colhead{$\Delta \lambda$($\mu$m)} &
\colhead{0.131} &
\colhead{0.163} &
\colhead{0.252} &
\colhead{0.264} &
\colhead{0.687} &
\colhead{0.872} &
\colhead{1.25} &
\colhead{2.55} &
\colhead{5.51} &
\colhead{5.32} 
\\
\colhead{$A_\lambda/A_V$} &
\colhead{0.488} &
\colhead{0.296} &
\colhead{0.187} &
\colhead{0.116} &
\colhead{0.0451} &
\colhead{0.0288} &
\colhead{0.0193} &
\colhead{0.0296} &
\colhead{0.0355} &
\colhead{0.0193} 
}
\startdata
NGC0024&\nodata      &228\p012E$-$1&246\p013E$-$1&188\p012E$-$1&101\p013E$-$1&682\p101E$-$2&821\p116E$-$2&117\p016E$-$1&924\p064E$-$2&120\p004E$-$1\\
NGC0337&136\p002E$-$1&178\p009E$-$1&190\p009E$-$1&164\p008E$-$1&958\p126E$-$2&659\p090E$-$2&141\p018E$-$1&376\p047E$-$1&291\p020E$-$1&673\p028E$-$1\\
NGC0584&562\p011E$-$1&880\p044E$-$1&109\p005E$+$0&860\p043E$-$1&365\p049E$-$1&219\p029E$-$1&174\p022E$-$1&114\p014E$-$1&700\p049E$-$2&480\p020E$-$2\\
NGC0628&123\p002E$+$0&156\p007E$+$0&160\p008E$+$0&128\p006E$+$0&839\p117E$-$1&544\p075E$-$1&116\p014E$+$0&259\p033E$+$0&235\p016E$+$0&323\p012E$+$0\\
NGC0855&766\p017E$-$2&853\p051E$-$2&940\p064E$-$2&778\p066E$-$2&428\p060E$-$2&280\p039E$-$2&359\p030E$-$2&474\p057E$-$2&354\p024E$-$2&851\p034E$-$2\\
NGC0925&\nodata      &561\p028E$-$1&621\p031E$-$1&500\p026E$-$1&307\p041E$-$1&210\p028E$-$1&349\p044E$-$1&606\p075E$-$1&545\p038E$-$1&943\p040E$-$1\\
NGC1097&\nodata      &234\p011E$+$0&270\p013E$+$0&226\p011E$+$0&123\p016E$+$0&800\p109E$-$1&146\p018E$+$0&317\p039E$+$0&298\p020E$+$0&661\p027E$+$0\\
NGC1266&\nodata      &112\p005E$-$1&120\p006E$-$1&116\p006E$-$1&538\p073E$-$2&414\p054E$-$2&542\p072E$-$2&878\p111E$-$2&111\p007E$-$1&872\p036E$-$1\\
\enddata     
\tablecomments{\footnotesize The compact table entry format TUV$\pm$WXYEZ implies (T.UV$\pm$W.XY)$\times10^{\rm Z}$ in Jy.  Corrections for neither Galactic nor intrinsic extinction has been applied (see \S~\ref{sec:data}).  The uncertainties include both statistical and systematic effects.  5$\sigma$ upper limits are provided for non-detections.  No color corrections have been applied.  The filter central wavelengths and widths are computed via Equations~\ref{eqn:filter_central} and \ref{eqn:filter_delta}.}
\tablenotetext{a}{\footnotesize The \Spitzer\ MIPS data for NGC~3034 suffer from saturation and thus the tabulated values are lower limits.}
\end{deluxetable}

\clearpage
\begin{deluxetable}{lrrrrrrrrrr}
\tablenum{2}
\def\a{\tablenotemark{a}}
\def\p{$\pm$}
\floattable
\rotate
\tabletypesize{\tiny}
\tablecaption{Global Flux Densities in Janskys Corrected for Neither Galactic Nor Intrinsic Extinction}
\tablewidth{0pc}
\tablehead{
\colhead{} &
\colhead{\Spitzer} &
\colhead{\Herschel} &
\colhead{\Herschel} &
\colhead{\Spitzer} &
\colhead{\Herschel} &
\colhead{\Herschel} &
\colhead{\Herschel} &
\colhead{\Herschel} &
\colhead{\JCMT} &
\colhead{\Planck} 
\\
\colhead{Filter} &
\colhead{MIPS} &
\colhead{PACS} &
\colhead{PACS} &
\colhead{MIPS} &
\colhead{PACS} &
\colhead{SPIRE} &
\colhead{SPIRE} &
\colhead{SPIRE} &
\colhead{SCUBA} &
\colhead{HFI} 
\\
\colhead{$\bar{\lambda}$($\mu$m)} &
\colhead{72.5} &
\colhead{71.8} &
\colhead{103} &
\colhead{157} &
\colhead{167} &
\colhead{252} &
\colhead{353} &
\colhead{511} &
\colhead{850} &
\colhead{850} 
\\
\colhead{$\Delta \lambda$($\mu$m)} &
\colhead{21.3} &
\colhead{21.3} &
\colhead{30.6} &
\colhead{35.8} &
\colhead{68.5} &
\colhead{67.3} &
\colhead{95.2} &
\colhead{184} &
\colhead{247} &
\colhead{238} 
\\
\colhead{$A_\lambda/A_V$} &
\colhead{0.00231} &
\colhead{0.00231} &
\colhead{0.00104} &
\colhead{0.000388} &
\colhead{0.000388} &
\colhead{0.000152} &
\colhead{0.0000725} &
\colhead{0.0000361} &
\colhead{0.0000146} &
\colhead{0.0} 
}
\startdata
NGC0024&222\p015E$+$0&\nodata      &\nodata      &712\p085E$+$0&\nodata      &\nodata      &\nodata      &\nodata      &\nodata      &457\p091E$-$1\\
NGC0337&111\p007E$+$1&139\p006E$+$1&210\p010E$+$1&200\p024E$+$1&197\p009E$+$1&894\p063E$+$0&405\p028E$+$0&158\p011E$+$0&349\p052E$-$1&573\p234E$-$1\\
NGC0584&175\p045E$-$1&180\p025E$-$1&300\p033E$-$1&118\p030E$+$0&610\p038E$-$1&$<$838E$-$1  &$<$603E$-$1  &$<$419E$-$1  &\nodata      &\nodata      \\
NGC0628&338\p023E$+$1&420\p021E$+$1&836\p041E$+$1&111\p013E$+$2&114\p005E$+$2&621\p044E$+$1&298\p021E$+$1&114\p008E$+$1&\nodata      &198\p015E$+$0\\
NGC0855&168\p011E$+$0&208\p010E$+$0&253\p013E$+$0&222\p026E$+$0&206\p010E$+$0&135\p009E$+$0&726\p052E$-$1&222\p018E$-$1&\nodata      &\nodata      \\
NGC0925&143\p010E$+$1&136\p006E$+$1&289\p014E$+$1&433\p052E$+$1&372\p018E$+$1&251\p017E$+$1&138\p009E$+$1&652\p046E$+$0&\nodata      &\nodata      \\
NGC1097&598\p046E$+$1&787\p039E$+$1&123\p006E$+$2&153\p018E$+$2&132\p006E$+$2&669\p047E$+$1&293\p020E$+$1&106\p007E$+$1&144\p078E$+$0&212\p012E$+$0\\
NGC1266&126\p009E$+$1&148\p007E$+$1&171\p008E$+$1&102\p012E$+$1&113\p005E$+$1&404\p028E$+$0&145\p010E$+$0&402\p030E$-$1&\nodata      &\nodata      \\
\enddata     
\tablecomments{\footnotesize The compact table entry format TUV$\pm$WXYEZ implies (T.UV$\pm$W.XY)$\times10^{\rm Z}$ in Jy.  Corrections for neither Galactic nor intrinsic extinction has been applied (see \S~\ref{sec:data}).  The uncertainties include both statistical and systematic effects.  5$\sigma$ upper limits are provided for non-detections.  No color corrections have been applied.  The filter central wavelengths and widths are computed via Equations~\ref{eqn:filter_central} and \ref{eqn:filter_delta}.}
\tablenotetext{a}{\footnotesize The \Spitzer\ MIPS data for NGC~3034 suffer from saturation and thus the tabulated values are lower limits.}
\end{deluxetable}

\clearpage
\begin{deluxetable}{lcccc}
\tablenum{3}

\def\p{$\pm$}
\tabletypesize{\scriptsize}
\tablecaption{\B\V\Rc\Ic\ Photometry/Calibration Source}
\tablewidth{0pc}
\tablehead{
\colhead{Galaxy} &
\colhead{\B} &
\colhead{\V} &
\colhead{\Rc} &
\colhead{\Ic} 
}
\startdata
NGC0024&PS1    &PS1    &PS1    &PS1    \\
NGC0337&PS1    &PS1    &PS1    &PS1    \\
NGC0584&PS1    &PS1    &PS1    &PS1    \\
NGC0628&PS1    &PS1    &PS1    &PS1    \\
NGC0855&RC3    &RC3    &C14    &\nodata\\
NGC0925&PS1    &PS1    &PS1    &PS1    \\
NGC1097&MM09   &MM09   &MM09   &MM09   \\
NGC1266&PS1    &PS1    &PS1    &PS1    \\
\enddata
\tablecomments{\footnotesize PS1: recalibration of \cite{dale07} photometry using Pan-STARRS1 (\S\ref{sec:optical}); D07: \cite{dale07}; MM09: \cite{munozmateos09b}; C14: \cite{cook14a}; T09: unpublished photometry from M. Pierce via the Extragalactic Distance Database \citep{tully09}; RC3: \cite{devaucouleurs91}.}
\label{tab:cal}
\end{deluxetable}

\clearpage
\begin{deluxetable}{lrrrrr}
\tablenum{4}
\def\a{\tablenotemark{a}}
\def\p{$\pm$}
\floattable
\rotate
\tabletypesize{\tiny}
\tablecaption{WISE \& VLA 20~cm Total Flux Densities in Janskys Corrected for Neither Galactic Nor Intrinsic Extinction}
\tablewidth{0pc}
\tablehead{
\colhead{} &
\colhead{\WISE} &
\colhead{\WISE} &
\colhead{\WISE} &
\colhead{\WISE} &
\colhead{NRAO} 
\\
\colhead{Filter} &
\colhead{W1} &
\colhead{W2} &
\colhead{W3} &
\colhead{W4} &
\colhead{VLA} 
\\
\colhead{$\bar{\lambda}$($\mu$m)} &
\colhead{3.40} &
\colhead{4.65} &
\colhead{12.8} &
\colhead{22.4} &
\colhead{20~cm} 
\\
\colhead{$\Delta \lambda$($\mu$m)} &
\colhead{0.663} &
\colhead{1.04} &
\colhead{5.51} &
\colhead{4.11} &
\colhead{} 
\\
\colhead{$A_\lambda/A_V$} &
\colhead{0.0523} &
\colhead{0.0287} &
\colhead{0.0285} &
\colhead{0.0225} & 
\colhead{0.0} 
}
\startdata
NGC0024&108\p007E$-$1&544\p039E$-$2&893\p064E$-$2&111\p009E$-$1&\nodata      \\
NGC0337&977\p069E$-$2&670\p048E$-$2&308\p022E$-$1&724\p051E$-$1&109\p010E$-$1\\
NGC0584&420\p029E$-$1&208\p015E$-$1&661\p107E$-$2&335\p059E$-$2&$<$500E$-$2  \\
NGC0628&842\p059E$-$1&512\p037E$-$1&232\p016E$+$0&326\p023E$+$0&172\p017E$-$1\\
NGC0855&441\p031E$-$2&245\p017E$-$2&379\p028E$-$2&731\p055E$-$2&490\p050E$-$3\\
NGC0925&302\p021E$-$1&176\p012E$-$1&516\p037E$-$1&856\p066E$-$1&460\p050E$-$2\\
NGC1097&126\p008E$+$0&766\p055E$-$1&293\p037E$+$0&685\p083E$+$0&414\p041E$-$1\\
NGC1266&557\p039E$-$2&398\p029E$-$2&935\p067E$-$2&725\p051E$-$1&116\p012E$-$1\\
\enddata     
\tablecomments{\footnotesize The compact table entry format TUV$\pm$WXYEZ implies (T.UV$\pm$W.XY)$\times10^{\rm Z}$ in Jy.  Corrections for neither Galactic nor intrinsic extinction has been applied.  The uncertainties include both statistical and systematic effects.  No color corrections have been applied.  The filter central wavelengths and widths are computed via Equations~\ref{eqn:filter_central} and \ref{eqn:filter_delta}.}
\label{tab:WISE}
\end{deluxetable}

\clearpage
\floattable
\rotate
\begin{deluxetable}{lrrrrrrrrrrrr}
\tablenum{5}
\def\a{\tablenotemark{a}}
\def\p{$\pm$}
\tabletypesize{\tiny}
\tablecaption{Aperture Photometry for Cases of Upper Limits in Table~\ref{tab:fluxes}}
\tablewidth{0pc}
\tablehead{
\colhead{} &
\colhead{\Spitzer} &
\colhead{\Spitzer} &
\colhead{\WISE} &
\colhead{\Spitzer} &
\colhead{\Spitzer} &
\colhead{\Herschel} &
\colhead{\Herschel} &
\colhead{\Spitzer} &
\colhead{\Herschel} &
\colhead{\Herschel} &
\colhead{\Herschel} &
\colhead{\Herschel} 
\\
\colhead{Filter} &
\colhead{IRAC} &
\colhead{IRAC} &
\colhead{W3} &
\colhead{MIPS} &
\colhead{MIPS} &
\colhead{PACS} &
\colhead{PACS} &
\colhead{MIPS} &
\colhead{PACS} &
\colhead{SPIRE} &
\colhead{SPIRE} &
\colhead{SPIRE} 
\\
\colhead{$\bar{\lambda}$($\mu$m)} &
\colhead{5.76} &
\colhead{7.96} &
\colhead{12.8} &
\colhead{23.8} &
\colhead{72.5} &
\colhead{71.8} &
\colhead{103} &
\colhead{157} &
\colhead{167} &
\colhead{252} &
\colhead{353} &
\colhead{511} 
}
\startdata
NGC0584&\nodata         &\nodata      &\nodata         &\nodata      &\nodata      &\nodata         &\nodata         &\nodata         &\nodata      &   281\p158E$-$1&   110\p106E$-$1&099\p684E$-$2\\
NGC1404&\nodata         &\nodata      &\nodata         &\nodata      &110\p080E$-$1&   330\p062E$-$1&$-$111\p009E$+$0&   130\p120E$-$1&630\p066E$-$1&$-$149\p030E$-$1&$-$139\p025E$-$1&295\p166E$-$2\\
M81dwA &$-$010\p520E$-$3&034\p640E$-$3&$-$420\p090E$-$4&050\p330E$-$3&029\p146E$-$1&   330\p090E$-$2&\nodata         &   020\p473E$-$1&\nodata      &\nodata         &\nodata         &\nodata      \\
DDO053 &\nodata         &\nodata      &\nodata         &\nodata      &\nodata      &\nodata         &\nodata         &\nodata         &\nodata      &\nodata         &\nodata         &980\p578E$-$3\\
HoIX   &$-$400\p200E$-$4&771\p100E$-$3&   526\p050E$-$3&128\p040E$-$3&068\p120E$-$2&\nodata         &\nodata         &$-$251\p260E$-$2&\nodata      &\nodata         &\nodata         &\nodata      \\
DDO154 &   560\p080E$-$3&180\p200E$-$4&   422\p035E$-$3&145\p050E$-$3&338\p140E$-$2&$-$270\p019E$-$1&   500\p029E$-$1&$-$009\p280E$-$2&600\p136E$-$2&   109\p011E$-$1&   693\p093E$-$2&294\p061E$-$2\\
DDO165 &\nodata         & \nodata     &   800\p280E$-$4&449\p060E$-$3&141\p020E$-$1&$-$240\p021E$-$1&   150\p022E$-$1&   206\p043E$-$1&400\p178E$-$2&   101\p133E$-$2&   100\p124E$-$2&501\p092E$-$2\\
\enddata     
\tablecomments{\footnotesize The compact table entry format TUV$\pm$WXYEZ implies (T.UV$\pm$W.XY)$\times10^{\rm Z}$ in Jy.  Corrections for neither Galactic nor intrinsic extinction has been applied (see \S~\ref{sec:data}).  The uncertainties include both statistical and systematic effects.  No color corrections have been applied.}
\label{tab:upperlimits}
\end{deluxetable}


\addtocounter{figure}{-1}
\begin{figure} 
 \plotone{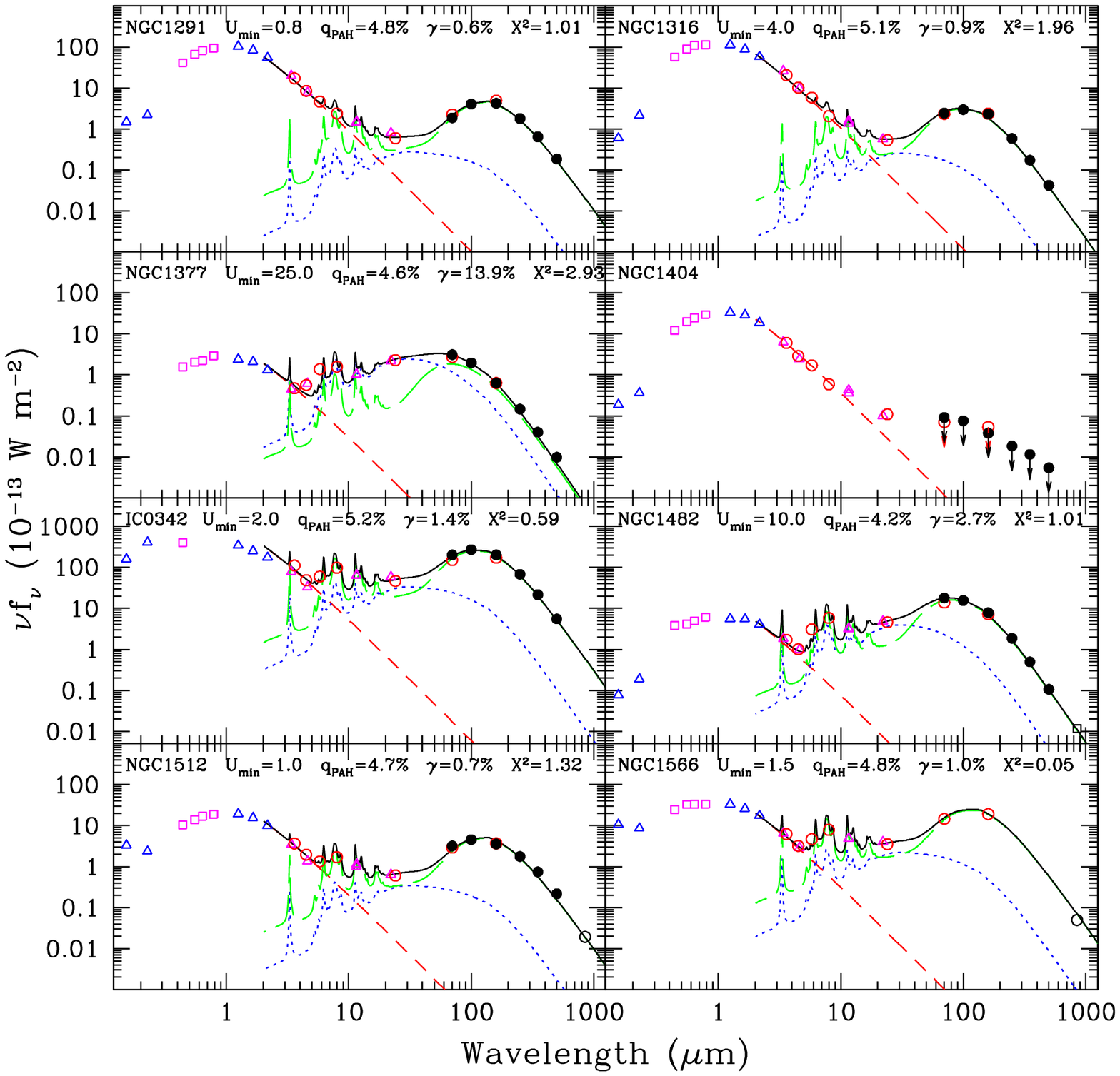} 
 \caption{Globally-integrated infrared/sub-millimeter spectral energy distributions for the KINGFISH/SINGS sample (continued).}
\end{figure}

\addtocounter{figure}{-1}
\begin{figure} 
 \plotone{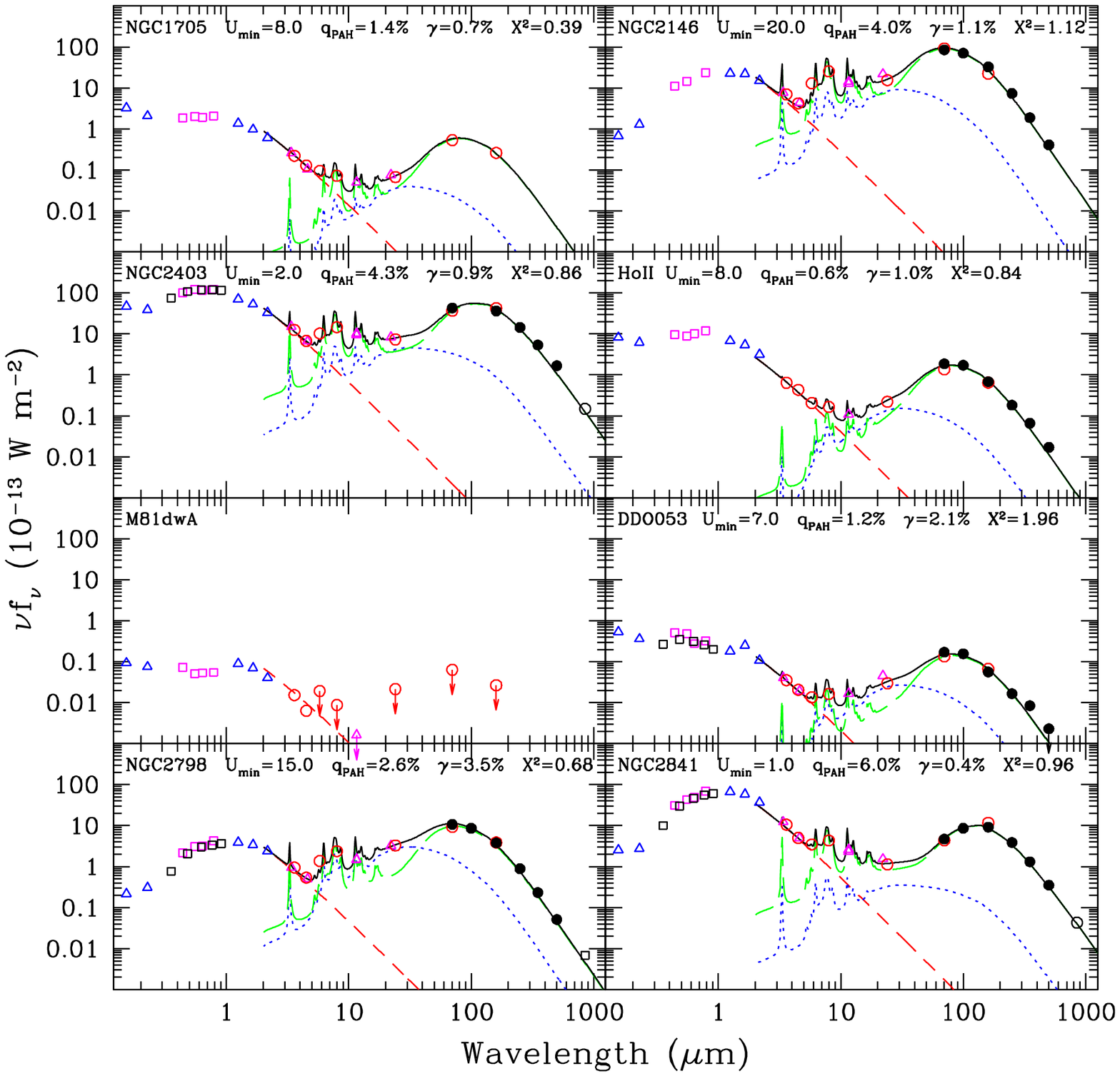} 
 \caption{Globally-integrated infrared/sub-millimeter spectral energy distributions for the KINGFISH/SINGS sample (continued).}
\end{figure}

\addtocounter{figure}{-1}
\begin{figure} 
 \plotone{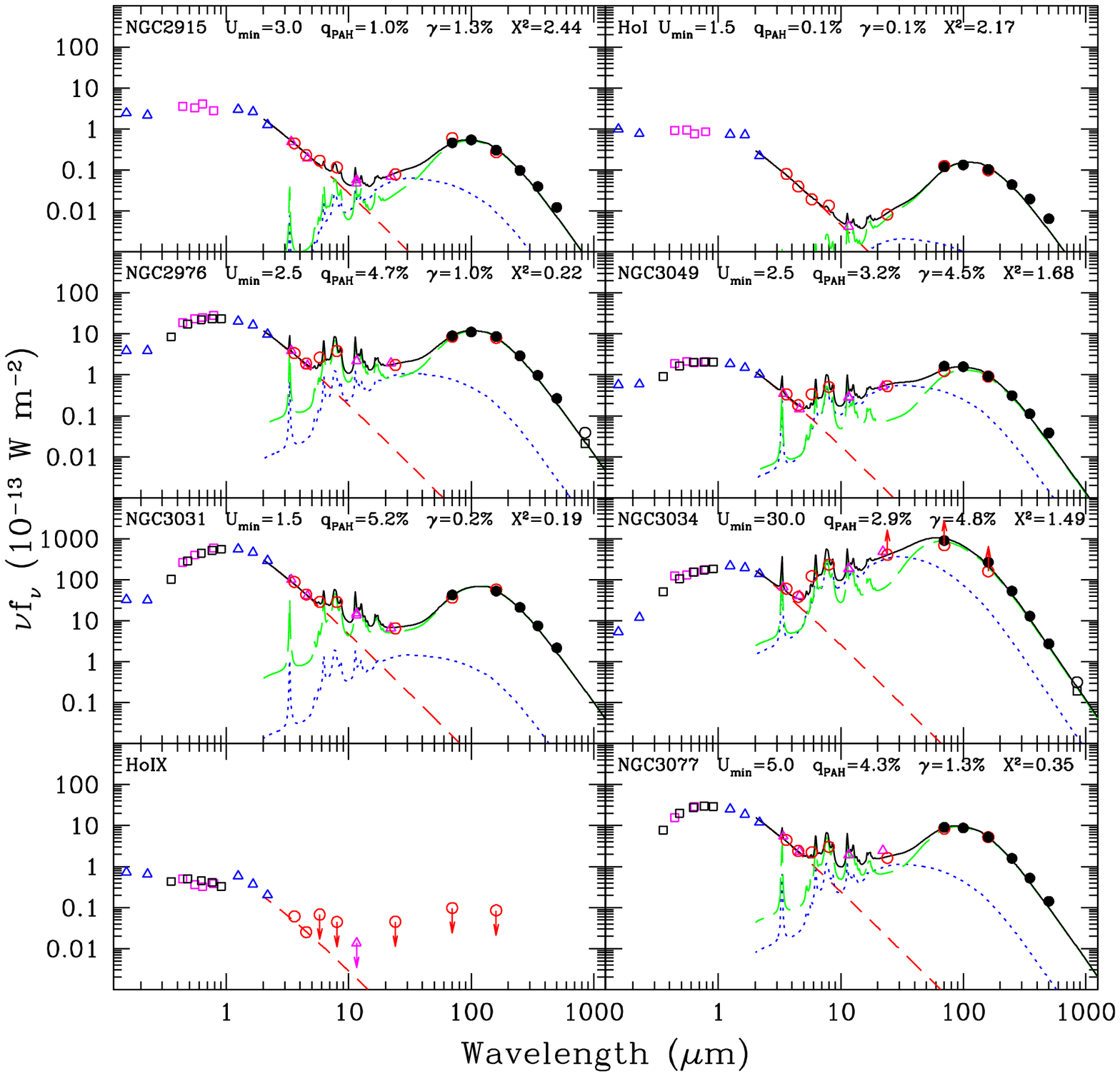} 
 \caption{Globally-integrated infrared/sub-millimeter spectral energy distributions for the KINGFISH/SINGS sample (continued).}
\end{figure}

\addtocounter{figure}{-1}
\begin{figure} 
 \plotone{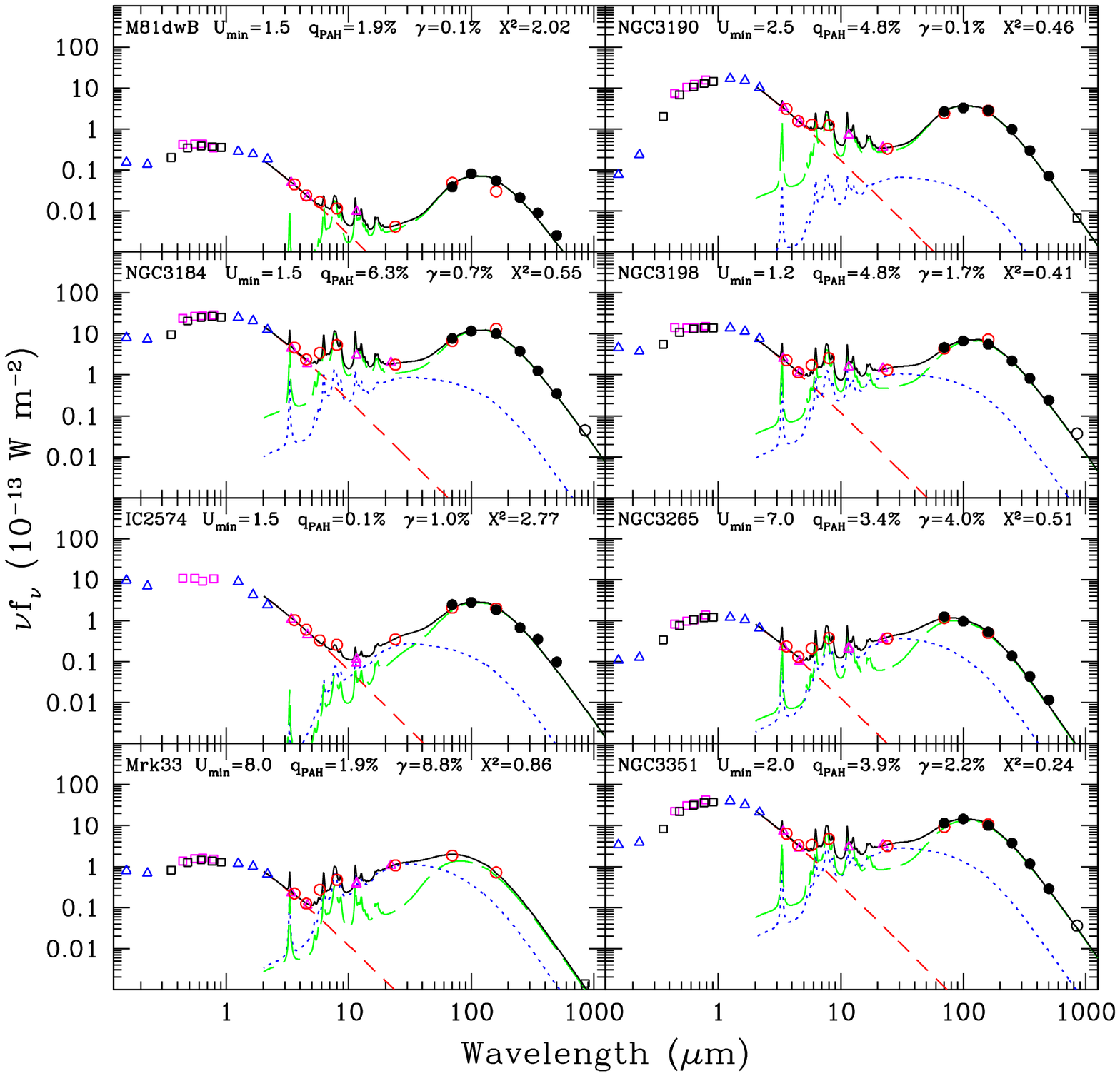} 
 \caption{Globally-integrated infrared/sub-millimeter spectral energy distributions for the KINGFISH/SINGS sample (continued).}
\end{figure}

\addtocounter{figure}{-1}
\begin{figure} 
 \plotone{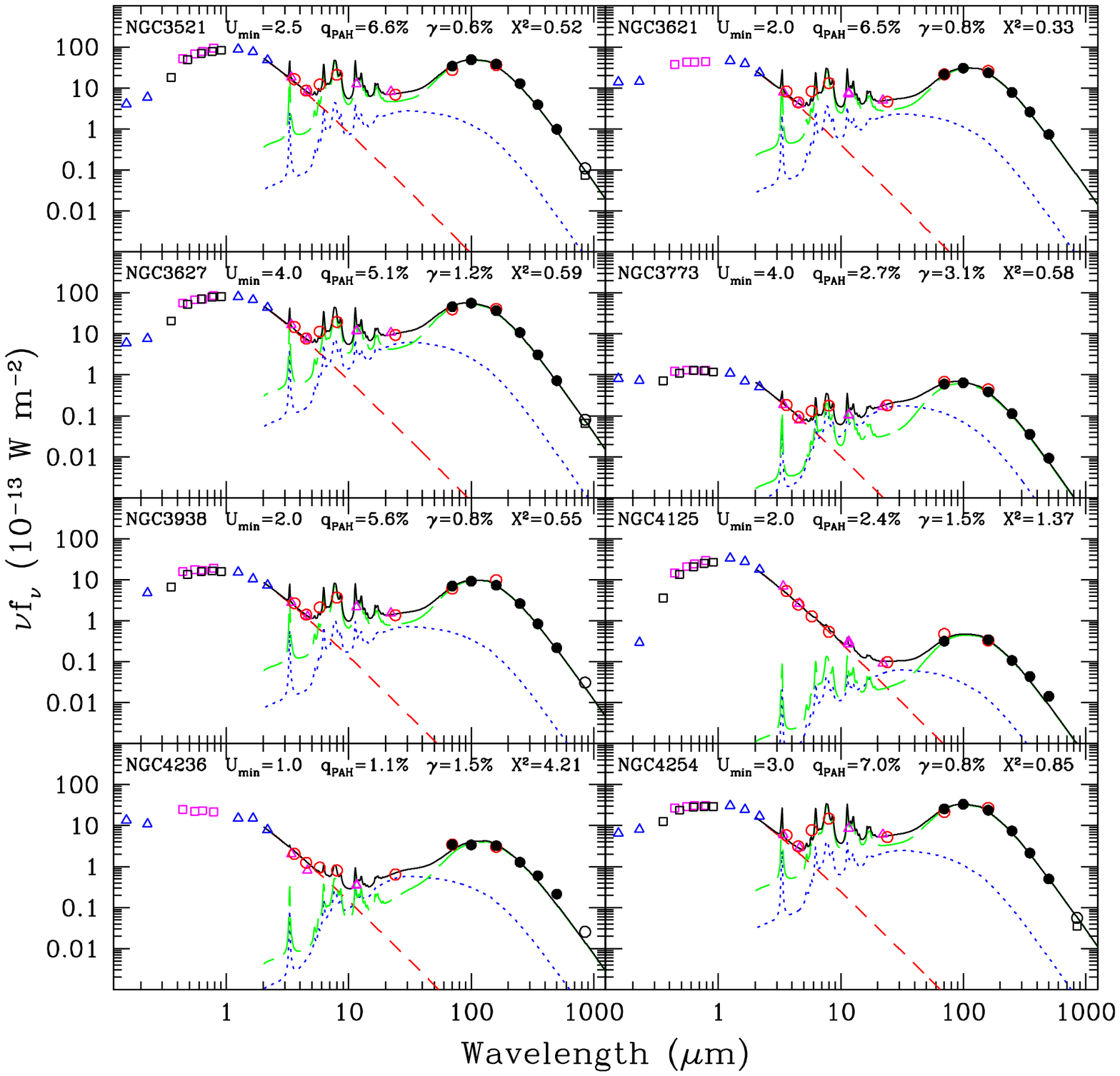} 
 \caption{Globally-integrated infrared/sub-millimeter spectral energy distributions for the KINGFISH/SINGS sample (continued).}
\end{figure}

\addtocounter{figure}{-1}
\begin{figure} 
 \plotone{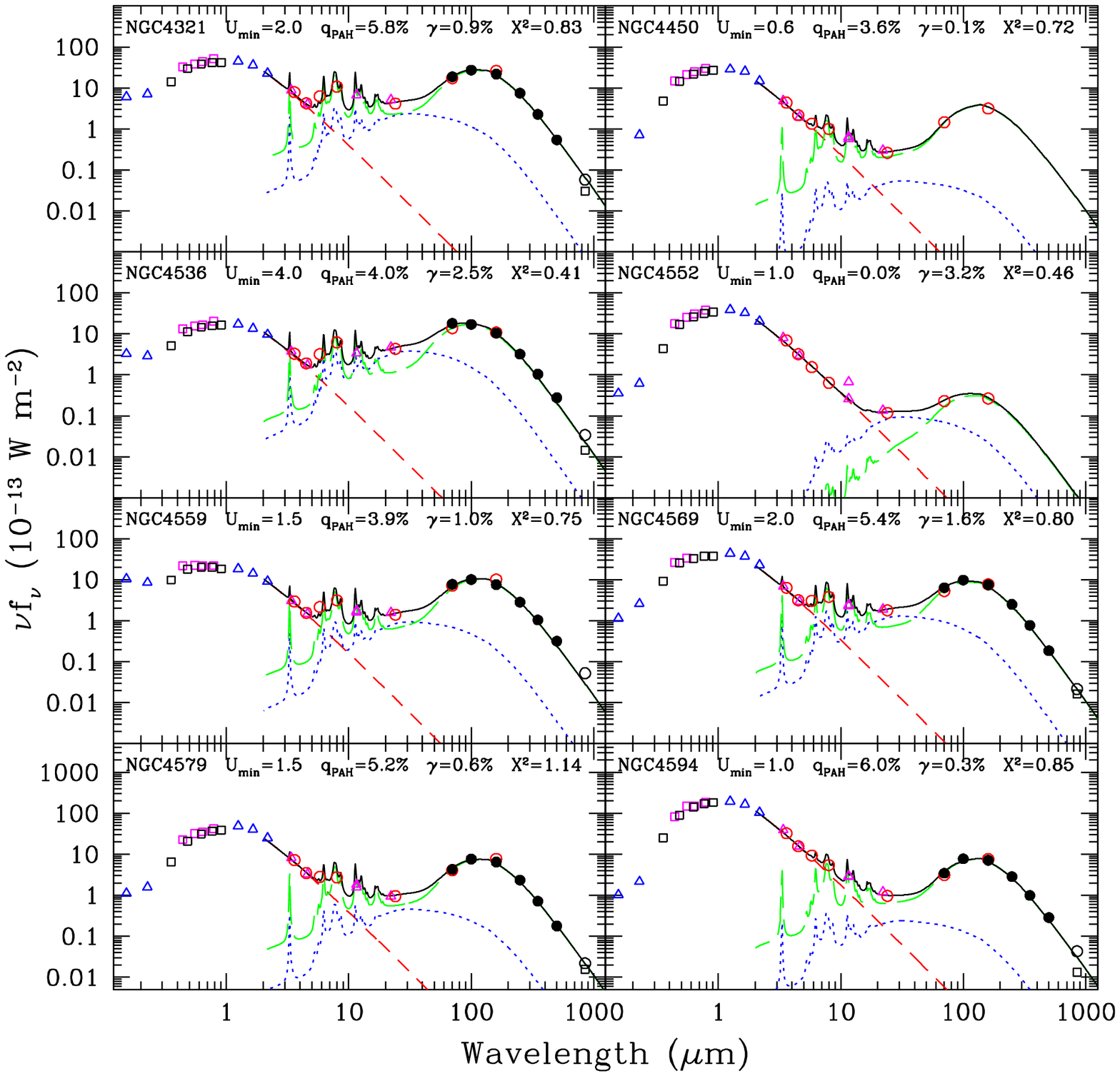} 
 \caption{Globally-integrated infrared/sub-millimeter spectral energy distributions for the KINGFISH/SINGS sample (continued).}
\end{figure}

\addtocounter{figure}{-1}
\begin{figure} 
 \plotone{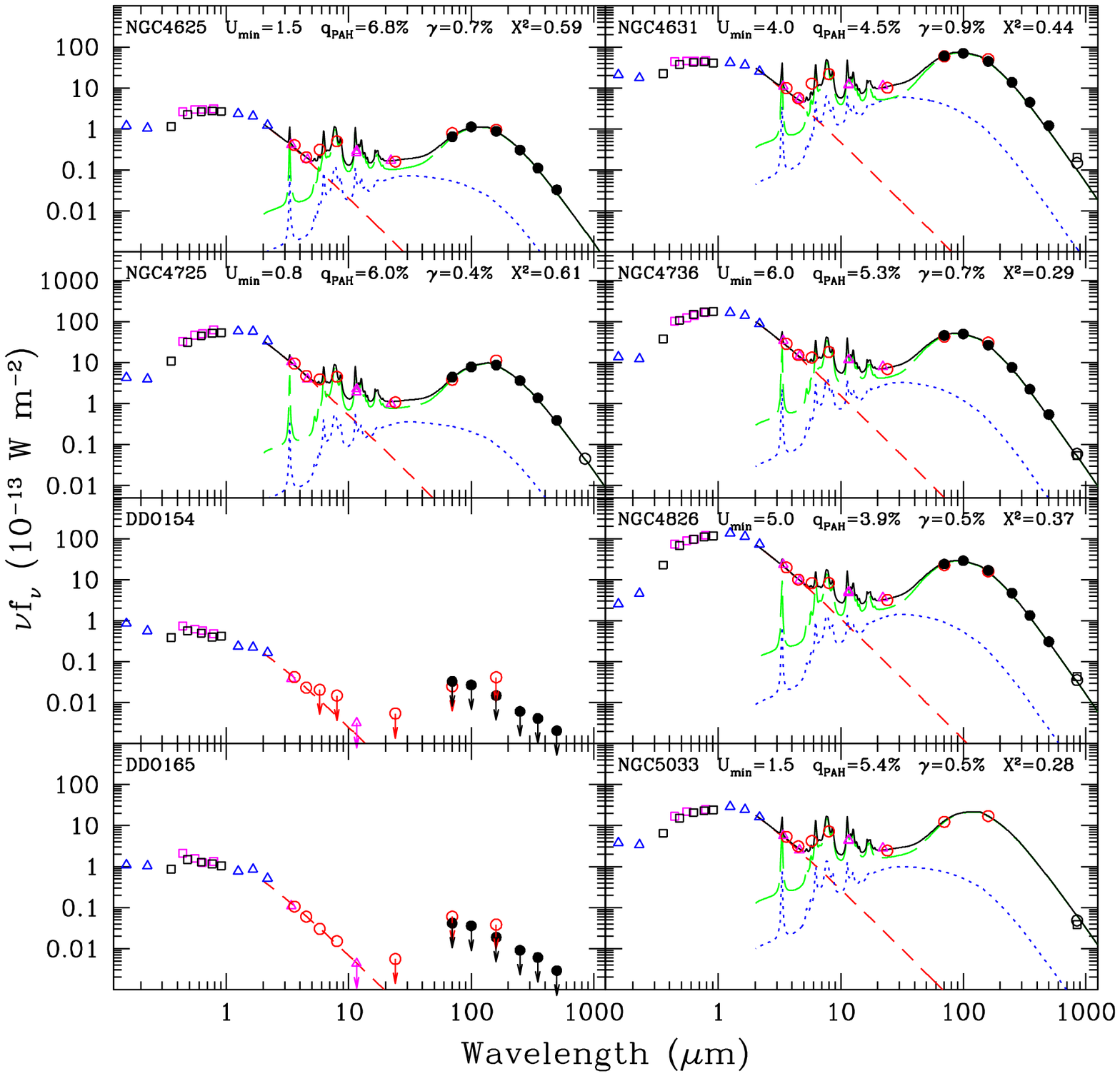} 
 \caption{Globally-integrated infrared/sub-millimeter spectral energy distributions for the KINGFISH/SINGS sample (continued).}
\end{figure}

\addtocounter{figure}{-1}
\begin{figure} 
 \plotone{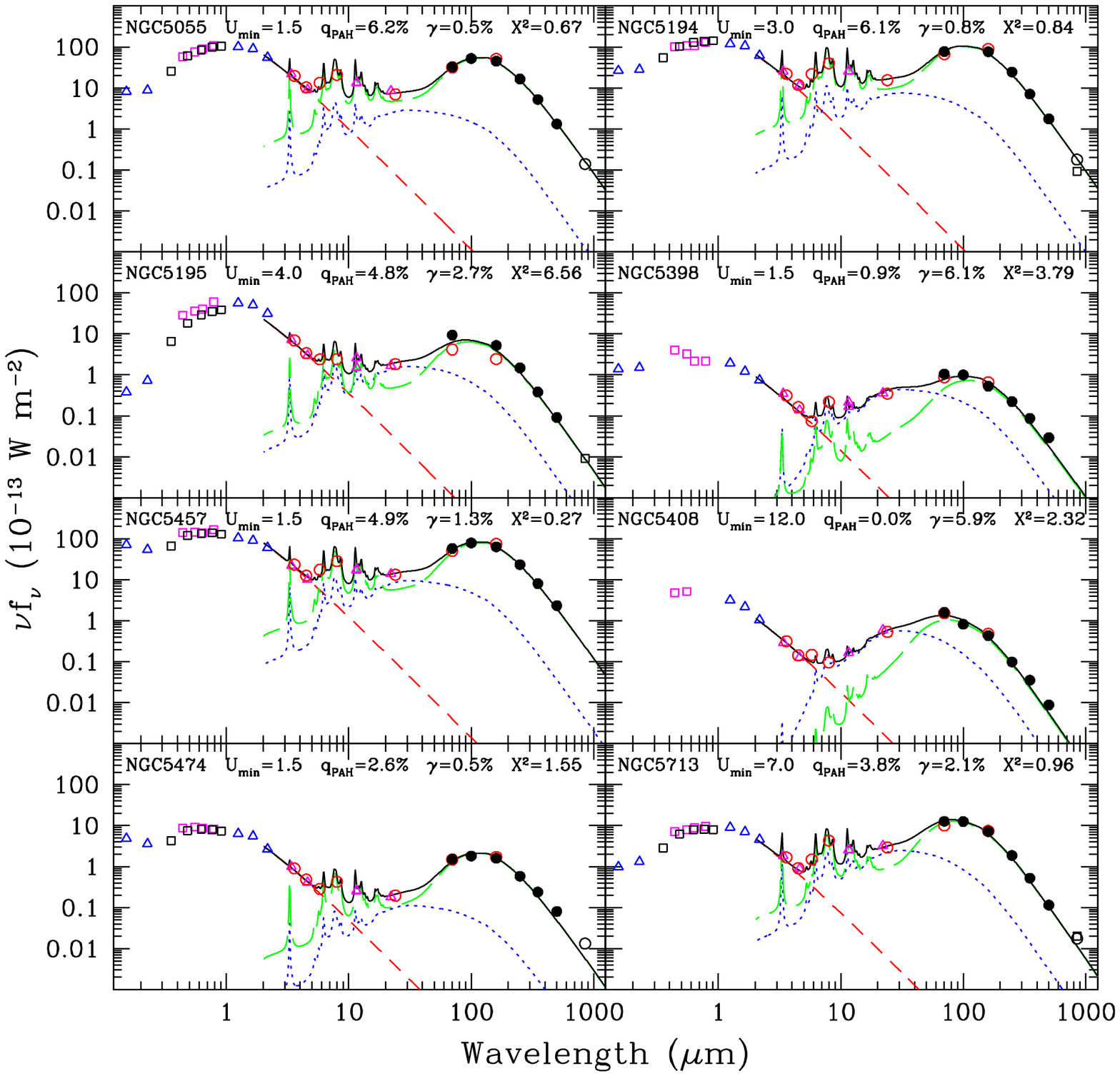} 
 \caption{Globally-integrated infrared/sub-millimeter spectral energy distributions for the KINGFISH/SINGS sample (continued).}
\end{figure}

\addtocounter{figure}{-1}
\begin{figure} 
 \plotone{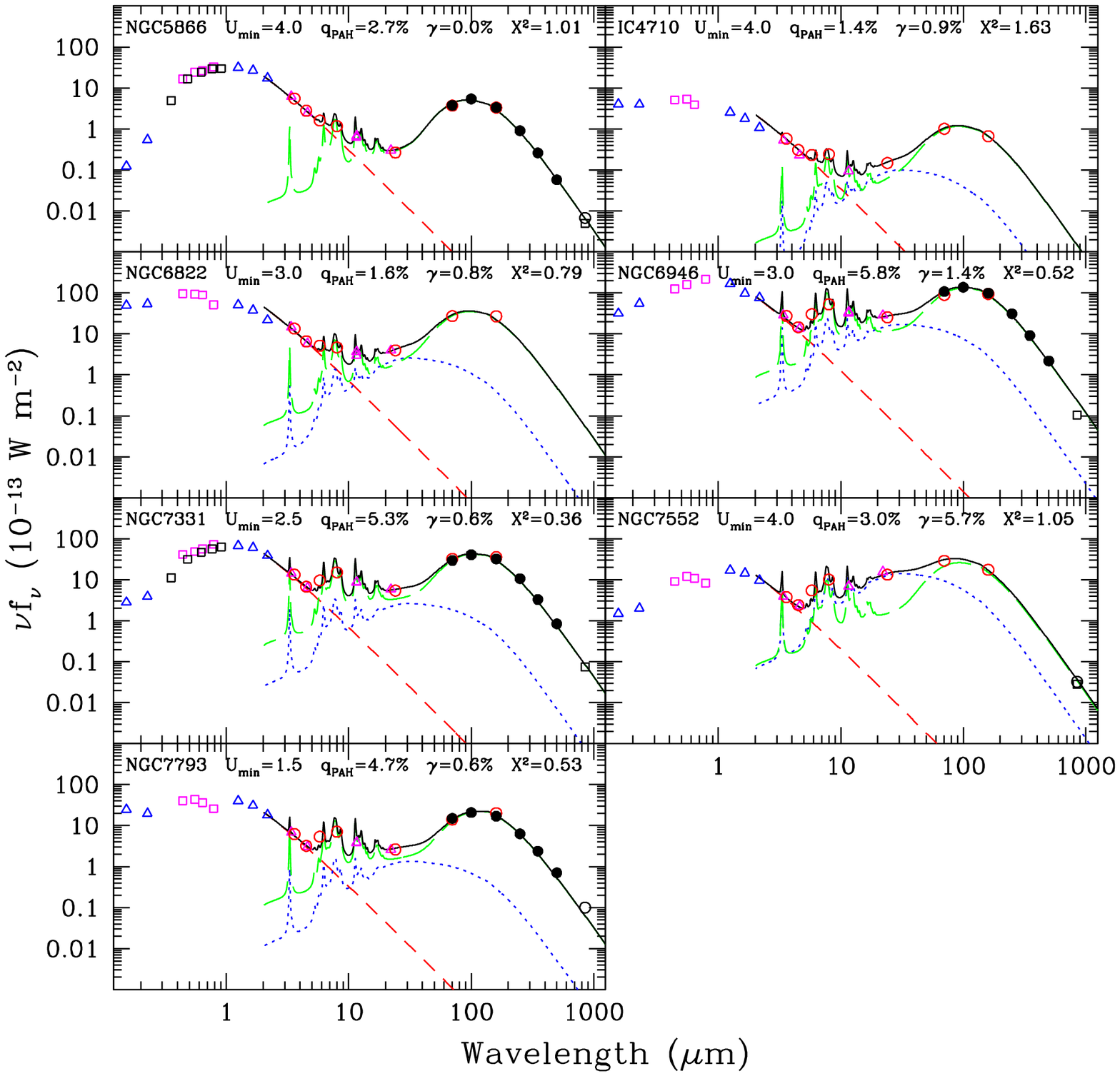} 
 \caption{Globally-integrated infrared/sub-millimeter spectral energy distributions for the KINGFISH/SINGS sample (continued).}
\end{figure}

\begin{figure}
 \plotone{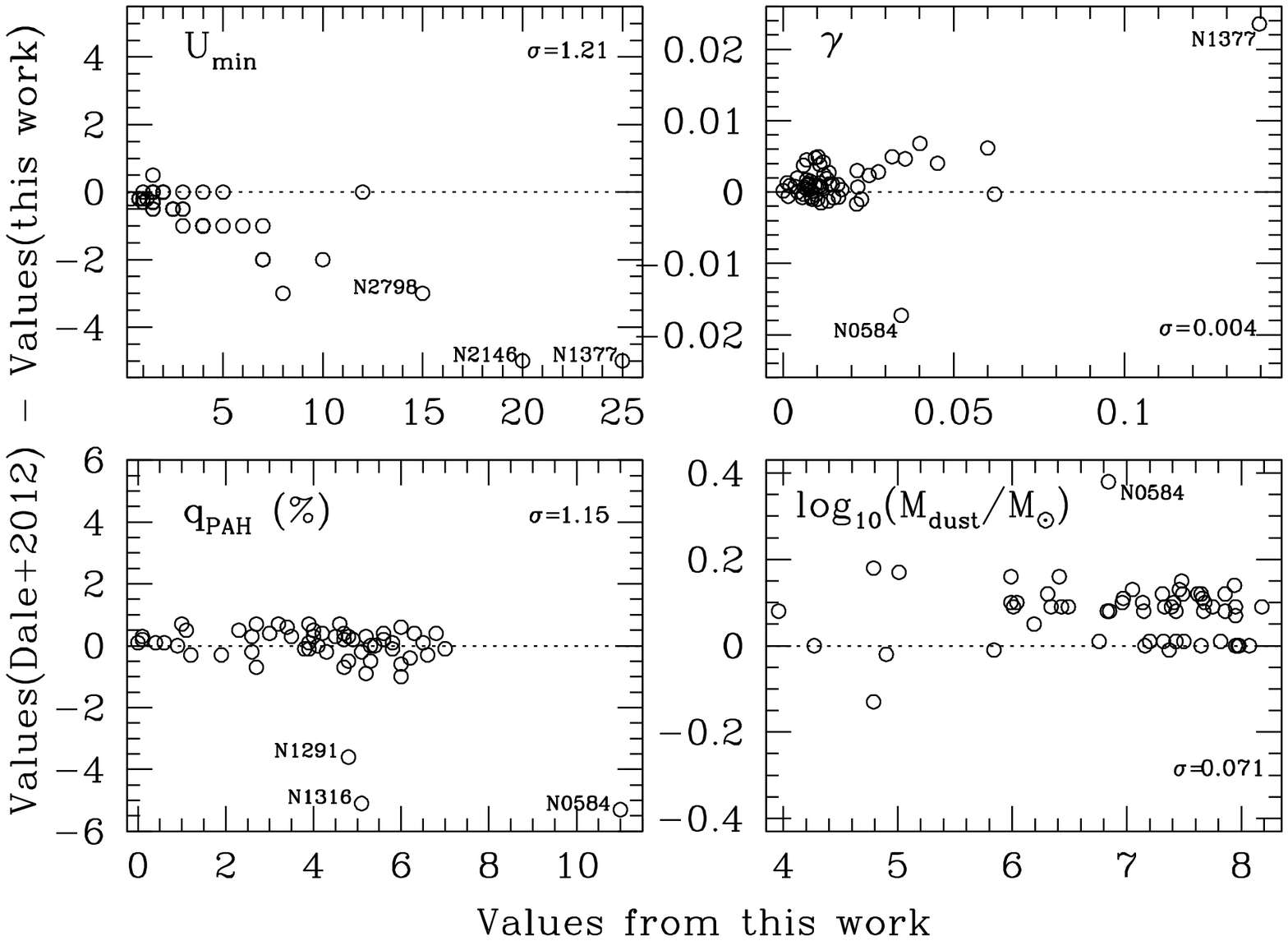}
 \caption{Comparison of DL07 output fit parameters from \cite{dale12} with those measured here.  The dispersions in the ordinate (y-axis) values are inset.}
 \label{fig:DL07}
\end{figure}

\begin{figure}
 \plotone{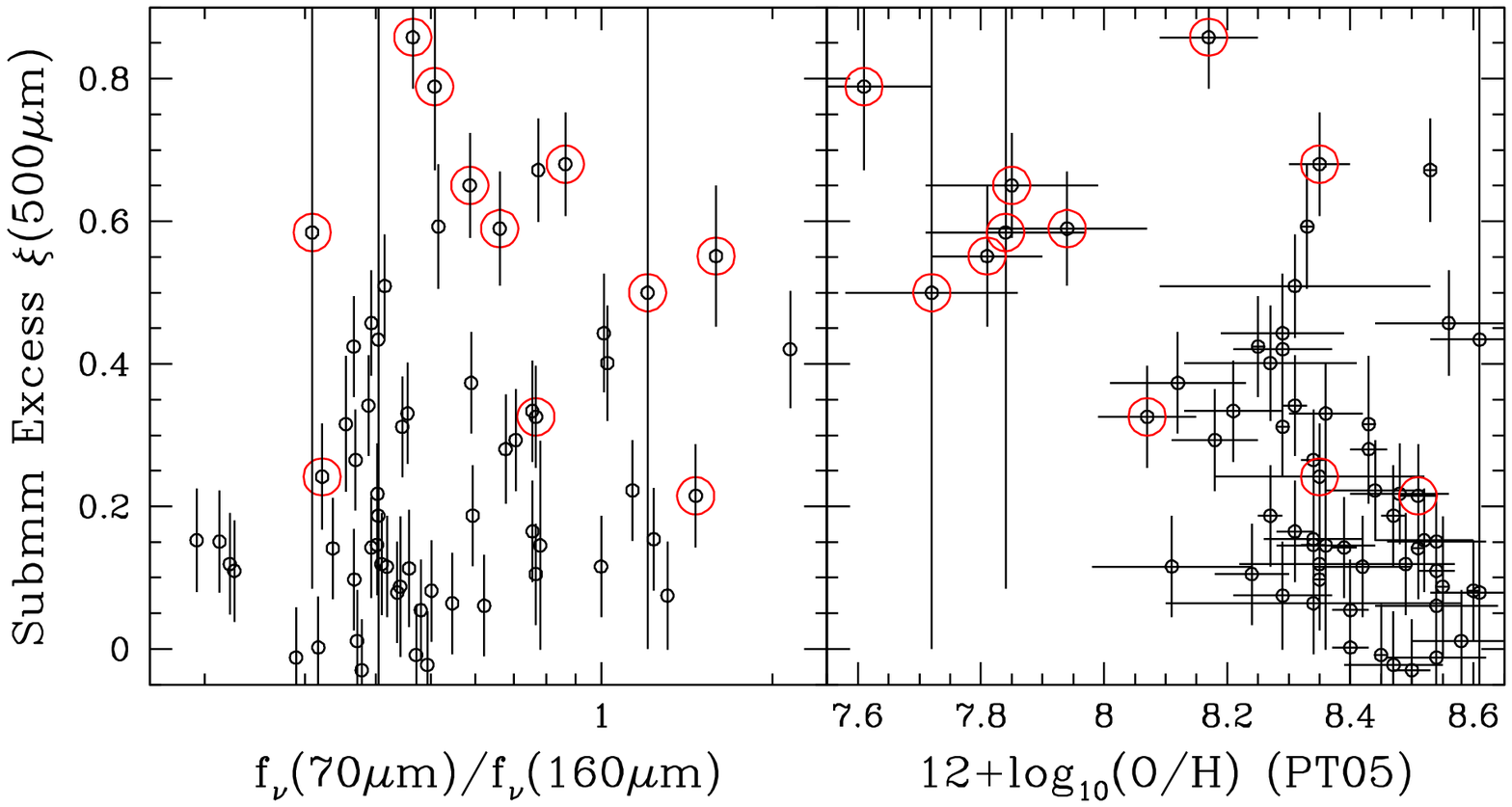}
 \caption{The submillimeter excess parameter $\xi(500\mu{\rm m})$ (see Equation~\ref{eq:excess}) as a function of far-infrared color and characteristic oxygen abundance as derived from \cite{moustakas10}.  Red circles indicate irregular galaxies ($T=$Sm, Im, or I0).}
 \label{fig:excess}
\end{figure}

\begin{figure}
 \plotone{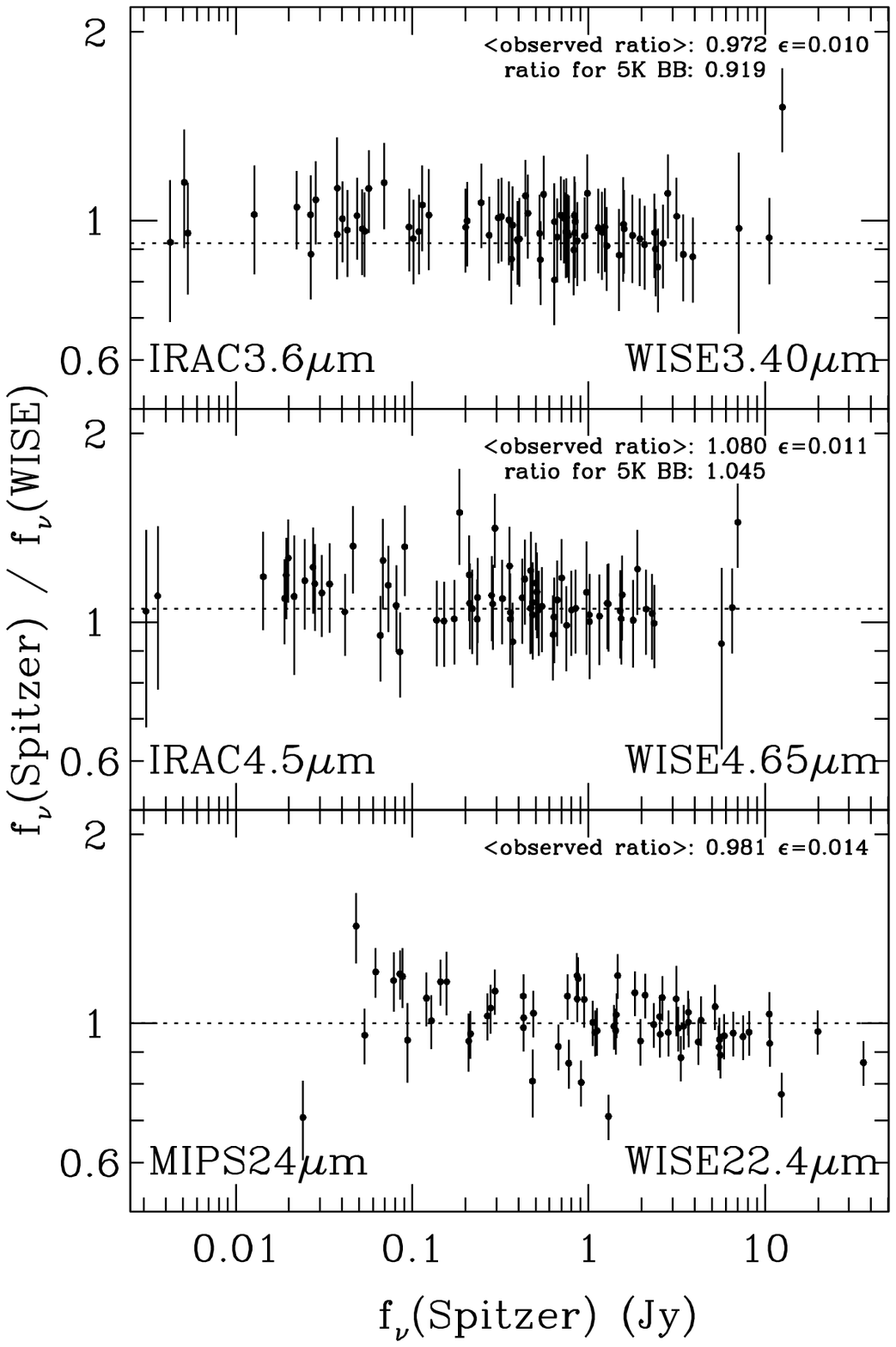}
 \caption{Comparison of \Spitzer- and aperture-based global photometry with total WISE photometry.  The dotted lines for the top two panels indicate the expected flux density ratios after convolving a 5000~K blackbody with the respective filter bandpass profiles.  The average ratio and its uncertainty ($\sigma/\sqrt{N}$) are inset.}
 \label{fig:wise}
\end{figure}

\end{document}